%% file: soda10.tex
\documentclass{article}

\usepackage{fullpage}
\usepackage{times}
\usepackage{epsfig}
\usepackage{amsmath}
\usepackage{amssymb}
\usepackage{amstext}
\usepackage{amsmath}
\usepackage{xspace}
\usepackage{latexsym}
\usepackage{verbatim}
\usepackage{multirow}

\newtheorem{theorem}{Theorem}
\newtheorem{definition}{Definition}

\newtheorem{lemma}[theorem]{Lemma}

\newtheorem{proposition}[theorem]{Proposition}
\newtheorem{corollary}[theorem]{Corollary}
\newtheorem{example}{Example}
\newcommand{\qed}{\mbox{\ \ \ }\rule{6pt}{7pt} \bigskip}

\renewcommand{\comment}[1]{}
\newenvironment{proof}{\noindent{\em Proof:}}{\hfill\qed}

\newenvironment{proofof}[1]{\noindent{\em Proof of #1.}}{}

\newcommand{\argmax}{\operatorname{argmax}}

\newcommand{\A}{{\mathcal A}}

\newcommand{\iid}{{i.~i.~d.\ }}

\newcommand{\val}{v}
\newcommand{\vals}{{\mathbf \val}}

\newcommand{\vali}[1][i]{{\val_{#1}}}

\newcommand{\virt}{\phi}

\newcommand{\virti}[1][i]{{\virt_{#1}}}

\newcommand{\ivirt}{\bar{\virt}}

\newcommand{\ivirti}[1][i]{{\ivirt_{#1}}}

\newcommand{\dist}{F}
\newcommand{\dists}{{\mathbf \dist}}
\newcommand{\disti}[1][i]{{\dist_{#1}}}

\newcommand{\dens}{f}

\newcommand{\densi}[1][i]{{\dens_{#1}}}

\newcommand{\price}{p}
\newcommand{\prices}{{\mathbf \price}}
\newcommand{\pricei}[1][i]{{\price_{#1}}}
\newcommand{\pricemi}[1][i]{{\price^M_{#1}}}

\newcommand{\rprice}{p}
\newcommand{\rprices}{{\mathbf \rprice}}
\newcommand{\rpricei}[1][i]{{\rprice_{#1}}}

\newcommand{\prob}{q}
\newcommand{\probs}{{\mathbf \prob}}
\newcommand{\probi}[1][i]{{\prob_{#1}}}
\newcommand{\probmi}[1][i]{{\prob^M_{#1}}}

\newcommand{\Mye}{\mathcal M}
\newcommand{\gMye}{{\mathcal M}^{G}}
\newcommand{\Alg}{\mathcal A}
\newcommand{\spm}[1][]{{\mathcal S}^{#1}}
\newcommand{\spmp}{{\mathcal S}'}
\newcommand{\gspm}{\spm[G]}
\newcommand{\vcg}[1][]{{\mathcal V}^{#1}}
\newcommand{\Algfc}{\Alg^{\text{copies}}}
\newcommand{\I}{\mathcal I}
\newcommand{\Ifc}{\I^{\text{copies}}}

\newcommand{\RevMye}{{\mathcal R}^{\Mye}}
\newcommand{\RevAlg}[1][\Alg]{{\mathcal R}^{#1}}
\newcommand{\Revp}[1][\prices]{{\mathcal R}^{#1}}
\newcommand{\RevSeq}[1][\prices]{{\mathcal R}^{\sigma}_{#1}}
\newcommand{\RevObl}[1][\prices]{{\mathcal R}^{\text{obl}}_{#1}}

\newcommand{\sets}{{\mathcal J}}
\newcommand{\rank}{\operatorname{rank}}
\newcommand{\spanOf}{\operatorname{span}}

\newcommand{\vsurp}{\Phi}

\newcommand{\expect}{\mathop{\operatorname{E}}}
\newcommand{\abs}[1]{\lvert{#1}\rvert}

\newcommand{\offerProb}[1][i]{c_{#1}}
\newcommand{\goodSet}{G}
\newcommand{\matroidElts}{E}

\newcommand{\gvirt}{{\virt^G}}
\newcommand{\gvirti}[1][i]{{\virt^G_{#1}}}
\newcommand{\probgi}[1][i]{{\prob^G_{#1}}}

\newcommand{\gi}{{\gamma_i}}
\newcommand{\gvs}{{\Phi^G}}

\newcommand{\soldset}{S}
\newcommand{\blockedset}{B}
\newcommand{\soldprice}[1]{p^{#1}}

\newcommand{\fixme}[1]{}

\begin{document}

\title{Multi-parameter Mechanism Design\\ and Sequential Posted Pricing}
\author{Shuchi Chawla\thanks{Computer Sciences Dept., University of Wisconsin -
  Madison. \tt{shuchi@cs.wisc.edu}.}
 \and Jason D. Hartline\thanks{EECS, Northwestern
   University. \tt{hartline@eecs.northwestern.edu}.}
 \and David Malec\thanks{Computer Sciences Dept., University of Wisconsin -
  Madison. \tt{dmalec@cs.wisc.edu}.}
 \and Balasubramanian Sivan\thanks{Computer Sciences Dept., University
 of Wisconsin - Madison. \tt{balu2901@cs.wisc.edu}.}
}
\date{}

\maketitle{}

\thispagestyle{empty}

\begin{abstract}
\input{abstract.tex}
\end{abstract}

\newpage
\setcounter{page}{1}

\section{Introduction}
\input{intro.tex}

\section{Problem set-up and preliminaries}
\input{prelim.tex}

\section{A reduction from multi-parameter MD to single-parameter MD}
\input{mpmd.tex}

\section{Sequential posted-price mechanisms}
\input{matroid.tex}

\section{Order-oblivious posted-prices}
\input{opm.tex}

\section{Approximations for the multi-parameter setting}
\input{mp-approx.tex}

\section{Discussion}
\input{discuss.tex}


\bibliographystyle{plain}
\bibliography{agt}

\appendix

\input{appendix.tex}

\end{document}

%% file: abstract.tex
We consider the classical mathematical economics problem of {\em
Bayesian optimal mechanism design} where a principal aims to optimize
expected revenue when allocating resources to self-interested agents
with preferences drawn from a known distribution.  In single-parameter
settings (i.e., where each agent's preference is given by a single
private value for being served and zero for not being served) this
problem is solved~\cite{mye-81}.  Unfortunately, these single
parameter optimal mechanisms are impractical and rarely
employed~\cite{ausubel06a}, and furthermore the underlying economic
theory fails to generalize to the important, relevant, and unsolved
multi-dimensional setting (i.e., where each agent's preference is
given by multiple values for each of the multiple services
available)~\cite{MV-07}.

In contrast to the theory of optimal mechanisms we develop a theory of
sequential posted price mechanisms, where agents in sequence are
offered take-it-or-leave-it prices.  We prove that these mechanisms
are approximately optimal in single-dimensional settings.  These
posted-price mechanisms avoid many of the properties of optimal
mechanisms that make the latter impractical.  Furthermore, these
mechanisms generalize naturally to multi-dimensional settings where
they give the first known approximations to the elusive optimal
multi-dimensional mechanism design problem.  In particular, we solve
multi-dimensional multi-unit auction problems and generalizations to
matroid feasibility constraints.  The constant approximations we
obtain range from~1.5 to~8.  For all but one case, our posted price
sequences can be computed in polynomial time.

This work can be viewed as an extension and improvement of the
single-agent algorithmic pricing work of~\cite{CHK-07} to the setting
of multiple agents where the designer has combinatorial feasibility
constraints on which agents can simultaneously obtain each service.

%% file: intro.tex




Suppose the local organizers for a prominent symposium on computer
science need to arrange for suitable hotel accommodations in the
Boston area for the attendees of the conference.  There are a number
of hotel rooms available with different features and attendees have
preferences over the rooms.  The organizers need a mechanism for
soliciting preferences, assigning rooms, and calculating
payments. Fortunately, they have distributional knowledge over the
participants' preferences (e.g., from similar conferences). This is a
stereotypical multi-dimensional setting for mechanism design that, for
instance, also arises in most resource allocation problems in the
Internet.  What mechanism should the organizers employ to maximize
their objective (e.g., revenue)?

The economic theory of optimal mechanism design is elegant and
predictive in single-dimensional settings.  Here Myerson's theory of
virtual valuations and characterizations of incentive constraints via
monotonicity guide the design of optimal truthful
mechanisms~\cite{mye-81} with practical (often non-truthful)
implementations~\cite{ausubel06a}.  The challenge of multi-dimensional
settings (e.g., in the likely case that conference attendees,
i.e., agents, have different values for different hotel rooms) is
two-fold.  First, multi-dimensional settings are unlikely to permit
succinct descriptions of optimal mechanisms~\cite{MM-88, RC98, MV-07}.
Second, optimal mechanisms in multi-dimensional settings are unlikely
to have practical implementations -- even asking agents to report
their true types across the many possible outcomes of the mechanism
may be impractical.  In summary, theory and practical considerations
from optimal mechanism design in single-dimensional settings fail to
generalize to multi-dimensional settings.


This paper approaches these issues through the lens of approximation.
Our main results are simple, practical, approximately optimal
mechanisms for a large class of multi-dimensional settings.  We
consider the multi-dimensional setting through a single dimensional
analogy wherein each multi-dimensional agent is represented by many
independent single-dimensional agents (e.g., one for each hotel room).
The optimal revenue for this single-dimensional setting is well
understood and, due to increased competition among agents,
upper-bounds that of the original multi-dimensional setting.  We
describe a ``sequential posted price'' mechanism for the
single-dimensional setting that is practical and approximately optimal
and, in contrast to the optimal single-dimensional mechanism, achieves
its approximation without inter-agent competition.  This gives a
robustness to deviations in modeling assumptions and, for instance,
the same mechanism continues to be approximately optimal in the
original multi-dimensional setting.  Therefore, our theory for
approximately optimal single-dimensional mechanisms generalizes to
multi-dimensional settings.

In the context of computer science literature this work is an
extension of {\em algorithmic pricing} (e.g., \cite{GHKKKM-05}) to
settings with multiple agents; it is unrelated to the standard
computational questions of {\em algorithmic mechanism design} (e.g.,
\cite{LOS-99,NR-99}).  The central problem in algorithmic pricing can
be viewed (for the most part) as Bayesian revenue maximization in a
single agent setting (e.g.,~\cite{GHKKKM-05}).  Algorithmic pricing is
hard to approximate when the agent's values for different outcomes are
generally correlated~\cite{bri-06}; however, when the values are
independent there is a 3-approximation~\cite{CHK-07}. In this context,
our results improve and extend the independent case to settings with
multiple agents and combinatorial feasibility constraints.  Notice
that the challenge in these problems is one imposed by the
multi-dimensional incentive constraints and not one from an inherent
complexity of an underlying non-game-theoretic optimization
problem. (E.g., in the hotel example the underlying optimization
problem is simply maximum weighted matching.)  In contrast, most work
in algorithmic mechanism design addresses settings where economic
incentives are well understood but the underlying optimization
problem is computationally intractable (e.g., combinatorial
auctions~\cite{LOS-99}).


While our exposition focuses on revenue maximization, all of our
techniques and results apply equally well to {\em social welfare}.
Social welfare is unique among objectives in that designing optimal
mechanisms in multi-dimensional settings is solved (by the VCG
mechanism).  Therefore, the interesting implication of our work on
social welfare maximization is that sequential posted pricing
approximates the welfare of the VCG mechanism and may be more
practical.

\paragraph*{Sequential Posted Pricing.}

Consider a single-parameter setting where each agent has a private
value for service and there is a combinatorial feasibility constraint
on the set of agents that can be simultaneously served.  For this
setting a {\em sequential posted pricing} (SPM) is a mechanism defined
by a price for each agent, a sequence on agents, and the semantics
that each agent is offered their corresponding price in sequence as a
take-it-or-leave-it while-supplies-last offer.  Meaning: if it is
possible to serve the agent given the set of agents already being
served then the agent is offered the price.  A rational agent will
accept if and only if the price is no more than their private value
for service.  That prices are associated with the agents and not the
sequence reflects the possibility that agents may play asymmetric
roles for a given feasibility constraint or value distribution.

Consider the following hotel rooms example with one room, two
attendees, and attendee values independently and identically distributed
uniformly between \$100 and \$200.  The optimal mechanism is the
Vickrey auction and its expected revenue is \$133.  The optimal
sequential posted pricing is for the organizers to offer the room to
attendee~1 at a price of \$150. If the attendee accepts,
then the room is sold, otherwise it is offered to attendee~2
for \$100. The expected revenue of this SPM is \$125.  

We are interested in comparing the optimal mechanism to the optimal
posted pricing in general settings.  A special class of SPMs is one
where mechanisms have provable performance guarantees for any sequence
of the agents. These {\em order-oblivious posted pricings} (OPM) are
mechanisms defined by a price for each agent and the semantics that
each agent is offered their corresponding price in some arbitrary
sequence as a take-it-or-leave-it while-supplies-last offer.

In single-dimensional settings, the advantages of sequential posted
pricings speak to the many reasons optimal auctions are rarely seen in
practice~\cite{ausubel06a}, and explain why posted pricings are
ubiquitous~\cite{ebay}.  First, take-it-or-leave-it offers result in
trivial game dynamics: truthful responding is a dominant strategy.
Second, SPMs satisfy strong notions of collusion resistance, e.g.,
{\em group strategyproofness}~(see~\cite{GH-05}): the only way in
which an agent can ``help'' another agent is to decline an offer that
he could have accepted, thereby hurting his own utility.  Third,
agents do not need to precisely know or report their value, they must
only be able to evaluate their offer; therefore, they risk minimal
exposure of their private information.  Fourth, agents learn
immediately whether they will be served or not.  In conclusion, the
robustness of SPMs in single-dimensional settings makes their
approximation of optimal mechanisms independently worthy of study.

The final robustness property of SPMs, which is of paramount
importance to our study of the multi-dimensional setting, is that they
minimize the role of agent competition which implies that
single-dimensional SPMs can be used ``as-is'' in multi-dimensional
settings with only a constant factor loss in performance.  In our
translation from the multi-dimensional setting to the
single-dimensional setting, each multi-dimensional agent has many
single-dimensional representatives.  A good OPM for the
single-dimensional setting can be viewed as an OPM for the
multi-dimensional setting by grouping all representatives of an agent
together and making their offers simultaneously to the agent.  The
agent will of course accept the offer that maximizes their utility.
The resulting mechanism is truthful and achieves the same performance
guarantee as the single-parameter OPM.  For SPMs where we are not free
to group each multi-dimensional agent's single-dimensional
representative together, an agent possibly faces a strategic dilemma
of whether to accept an offer (e.g., for one hotel room) early on or
wait for a later offer (e.g., another hotel room) which may or may not
still be available.  Our guarantee is that regardless of the actions
of any agent with such a strategic option (i.e., {\em implementation
  in undominated strategies}, see, e.g.,~\cite{BLP09}) our performance
is a constant fraction of the original SPM's performance.  Given the
advantages of SPMs over truthful mechanisms, such a non-truthful SPM
may be more practically relevant than a truthful implementation.

Finally, we note that most of our results for posted pricings are
constructive and give efficient algorithms for them.  A posted price
mechanism has two components where computation is necessary: an
offline computation of the prices to post (and for SPMs, the sequence
of agents) and an online while-supplies-last offering of said
prices.\footnote{This is similar, for example, to nearest neighbor
  algorithms, where one distinguishes the time taken to construct a
  database, and the time taken to compute nearest neighbors over that
  database given a query.}  The agents are only present for the online
part where the mechanism is trivial.  All of the computational burden
for an SPM is in the offline part.  The offline computation of our
posted price mechanisms is based on a subroutine that repeatedly
samples the distribution of agent values and simulates Myerson's
mechanism on the sample.  This clearly requires more computation than
just running Myerson's mechanism on the real agents in the first
place; however, we benefit from the robustness that comes from the
trivial online implementation of posted pricings.

\paragraph{Related work.}

See \cite{MV-07} and references therein for work in economics on
optimal multi-dimensional mechanism design.  See \cite{CHK-07} and
references therein for work in computer science on multi-dimensional
pricing for a single agent.  We extend the setting from \cite{CHK-07}
to multiple agents and improve their approximation for a single agent
from $3$ to $2$.

Sequential posted price mechanisms have been considered previously in
single-dimensional settings.  Sandholm and Gilpin~\cite{SG06} show
experimentally that these mechanisms compare favorably to Myerson's
optimal mechanisms.  Blumrosen and Holenstein~\cite{BH-08} show how to
compute the optimal posted prices in the special case where agents'
values are distributed identically, and also show that in this case
the revenue of these mechanisms approaches the optimal revenue
asymptotically. Several papers study revenue maximization through
online posted pricings in the context of adversarial values, albeit in
the simpler context of multi-unit auctions \cite{BKRW-03, KL-03,
  BH-05}.

The question of whether simple mechanisms can achieve near-optimal
revenue was considered recently by Hartline and
Roughgarden~\cite{HR-09}.  Except for their result on single-item
auctions with anonymous reserve prices, their VCG based mechanisms are
likely to suffer the same impracticality criticisms as the optimal
mechanism.  The essay ``The Lovely but Lonely Vickrey Auction'' by
Ausubel and Milgrom \cite{ausubel06a} discusses why this is the case.
As a consequence of the near-optimality of sequential posted prices,
we answer one of their open questions in the positive, namely, that
the gap between the revenue optimal mechanism and a VCG mechanism with
appropriate reserve prices is a constant (i.e., $2$) in matroid
settings but with arbitrary valuation distributions. This bound
matches their result for regular distributions.

Our setting of sequential posted pricing with a matroid constraint is
very closely related to the so-called matroid secretary
problem~\cite{BIK-07,BDGIT09,KP09}, but there are two important
differences: (a) they assume that agents' values are adversarial,
whereas in our setting they are drawn from known distributions, and
(b) in their setting agents arrive in random order, whereas we
consider optimized and adversarial orderings.  Some of our results are
reminiscent of that work, but our techniques are necessarily different.

Finally, our results for OPMs in the multi-unit auction setting are
based on work on prophet inequalities from optimal stopping
theory. While that work applies directly to the analysis of OPMs in
the single-item auction setting, we show that it extends to $k$-unit
auctions with no loss in approximation factor.

\paragraph{Our results.}
Our results are summarized in Tables~\ref{tab:results-single} and
\ref{tab:results-multi}. Our approximation factors in both the
single-dimensional and multi-dimensional settings depend on the kind
of feasibility constraint that the seller faces. In the
single-dimensional setting, the feasibility constraint is a set system
over agents specifying the sets of agents that can be simultaneously
served. In the multi-dimensional setting, each agent is interested in
buying one of multiple kinds of items or services and we assume that
agents' values for the different services are independent. The
feasibility constraint is a set system over (agent, service) pairs. In
both cases we assume that the set system is downward closed, i.~e.,
any subset of a feasible set is also feasible. All of the mechanisms
we develop can be computed efficiently, except for the $O(\log k)$
approximate OPM in general matroid settings.

\begin{table*}[t]
\begin{small}
\begin{center}
\begin{tabular}{|l|l|c|c|l|}
\hline
\multirow{2}{*}{Feasibility constraint} & \multirow{2}{*}{Type of
mechanism} & \multicolumn{2}{|c|}{Gap from optimal} & \multirow{2}{*}{Reference}\\
& & upper bound & lower bound & \\
\hline
\hline
\multirow{2}{*}{General matroid} & SPM & 2 & $\sqrt{\pi/2}\approx 1.25$ & (\S~\ref{sec:spm-matroid}, \S~\ref{app:spm-bad-ex}, \cite{BH-08})\\
& OPM & $O(\log k)$ & 2 & (\S~\ref{subsection:opmMatroid}, \S~\ref{app:opm-uniform})\\
& VCG & 2 & - & (\S~\ref{sec:vcg})\\
\hline
\multirow{2}{*}{$k$-uniform matroid, partition matroid} & SPM & $e/(e-1)\approx 1.58$ &
$1.25$ & (\S~\ref{sec:spm-other}, \S~\ref{sec:spm-uniform})\\
& OPM & 2 & 2 & (\S~\ref{subsection:opmSpecialMatroid}, \S~\ref{app:opm-uniform})\\
\hline
Graphical matroid & OPM & 3 & 2 & (\S~\ref{subsection:opmSpecialMatroid}, \S~\ref{app:graphicalMatroid})\\
\hline
Intersection of two matroids & SPM & 3 & $1.25$ & (\S~\ref{sec:spm-other}, \S~\ref{app:spm-intersection})\\
\hline
Intersection of two partition matroids & OPM & 6.75 & 2 & (\S~\ref{subsection:opmMatroidIntersection}, \S~\ref{app:opm-intersection})\\
\hline
Non-matroid downward closed & SPM, OPM & - & $\Omega(\log n/\log\log n)$ & (\S~\ref{app:spm-nonmatroid})\\
\hline
\end{tabular}
\caption{A summary of our results for single-dimensional
preferences. Here $n$ is the number of agents, and $k$ is the size of the largest feasible set.}
\label{tab:results-single}
\end{center}
\end{small}
\end{table*}

\begin{table*}[t]
\begin{small}
\begin{center}
\begin{tabular}{|l|l|l|c|}
\hline
Feasibility constraint & Solution concept & Mechanism & Gap from optimal\\
\hline
\hline
multi-unit multi-item with unit-demand & dominant strategy truthful & OPM & 6.75\\
\hline
Graphical matroid with unit-demand & dominant strategy truthful & OPM & 32/3\\
\hline
General matroid intersection & alg. imp. in undominated strategies & SPM & 8\\
\hline
Combinatorial auction with small bundles & alg. imp. in undominated strategies & SPM & 8\\
\hline
\end{tabular}
\caption{A summary of our results for multi-dimensional
preferences (\S~\ref{sec:mp-approx}).}
\label{tab:results-multi}
\end{center}
\end{small}
\end{table*}

%% file: prelim.tex
\label{sec:prelim}
\subsection{Bayesian optimal mechanism design}
\label{subsec:BOMD}
In the single-parameter setting, the mechanism design problem we study
(hereafter abbreviated BSMD for Bayesian single-parameter mechanism
design) is stated as follows. There are $n$ single-parameter agents
and a single seller providing a certain service. Agent $i$'s value
$\vali$ for getting served is distributed independently according to
distribution function $\disti$ with density $\densi$. The seller faces
a feasibility constraint specified by a set system $\sets\subseteq
2^{[n]}$, and is allowed to serve any set of agents in $\sets$. We
assume that the set system $\sets$ is downward closed. That is, for
any $A\subset B\subseteq 2^{[n]}$, $B\in\sets$ implies $A\in\sets$. A
mechanism $M$ for this problem is a function that maps a vector of
values $\vals$ to an {\em allocation} $M(\vals)\in \sets$ and a {\em
  pricing} $\pi(\vals)$ with a price $\pi_i$ to be paid by agent $i$.

In the Bayesian multi-parameter unit-demand setting (BMUMD for short),
we again have $n$ buyers and one seller. The seller offers a number of
different services indexed by set $J$. The set $J$ is partitioned into
groups $J_i$, with the services in $J_i$ being targeted at agent
$i$.\footnote{Since we allow for an arbitrary feasibility constraint
over the set $J$, the assumption that the sets $J_i$ are disjoint is
without loss of generality.} Each agent $i$ is interested in getting
any one of the services in $J_i$ (that is, consumers are unit-demand
agents). In the hotel rooms example, the set $J_i$ would contain all
the rooms that agent $i$ may be interested in and the feasibility
constraint ensures that each room is allocated to at most one
agent. Another setting with a more general feasibility constraint
arises in the context of airline ticket sales: we have a directed
graph with capacities on edges owned by a seller, and a number of
agents. Each agent is interested in a path of at most two hops from
some source to some destination in the graph (agents want to buy
airline tickets for an itinerary with at most two legs), and $J_i$
contains all such paths. The feasibility constraint ensures that each
leg is allocated upto its capacity and no more.

Agent $i$ has value $\vali[j]$ for service $j\in J_i$. $\vali[j]$ is
independent of all other values and is drawn from distribution
$\disti[j]$. Once again the seller faces a feasibility constraint
specified by a set system $\sets\subseteq 2^{J}$. Note that for every
$S\in\sets$ and $i\in [n]$, $|S\cap J_i|\le 1$, that is each agent
gets at most one service. As in the single-parameter case, a mechanism
for this problem maps any set of bids $\vals$ to an allocation
$M(\vals)\in \sets$ and a pricing $\pi(\vals)$.


\subsection{Posted-price mechanisms}
\label{subsection:postedPriceMech}
A sequential posted-price mechanism (SPM), $\spm$, is defined by an
ordering $\sigma$ over agents and a collection of prices $\pricei$ for
$i\in [n]$. The mechanism is run as follows:
\begin{enumerate}
\item Initialize $A\leftarrow\emptyset$.
\item For $i=1$ through $n$, do:
\begin{enumerate}
 \item If $A\cup \{\sigma(i)\}\in\sets$, offer to serve agent
 $\sigma(i)$ at price $\pricei$.
 \item If the agent accepts, $A\leftarrow A\cup \{\sigma(i)\}$.
\end{enumerate}
\item Serve the agents in $A$.
\end{enumerate}

Let $c_i$ denote the probability taken over values of agents
$\sigma(1), \cdots, \sigma(i-1)$ that the mechanism offers to serve
agent $i$, and let $q_i=1-\disti(\pricei)$. Then the expected revenue
of the sequential mechanism, $\RevSeq$, is given by $\sum_i
c_iq_i\pricei$.

\paragraph{Order-oblivious posted prices.} As mentioned earlier, we
also study posted-price mechanisms where the order of offers is picked
adversarially. We estimate (pessimistically) the expected revenue of
this mechanism as follows:
\[\RevObl = \expect\nolimits_{\vals\sim\dists}{\min_{S\in {\mathcal S}_v} \sum_{i\in S}\pricei}\]
Here the minimization is over the class ${\mathcal S}_v$ of sets $S$
  that are {\bf maximal feasible subsets} of agents that ``desire''
  service given values $\vals$ and prices $\prices$: (1) ${\mathcal
  S}_v\subseteq\sets$, (2) $\vali\ge\pricei$ for all $i\in S$,
  $S\in {\mathcal S}_v$, and, (3) for every feasible superset $S'$ of a
  set $S\in {\mathcal S}_v$, $S'$ contains some agent $i$ with
  $\vali<\pricei$.

For some instances, we allow a strengthening of OPMs to posted-price
mechanisms where the seller is allowed to deny service to an agent
even when the agent can be feasibly served alongside previously served
agents. Formally, the mechanism selects a pricing $\prices$, and a set
system $\sets'\subseteq \sets$, and runs the OPM using the prices
$\prices$ but determining feasibility according to $\sets'$ instead of
$\sets$. Crucially, the system $\sets'$ is determined based only on
the distributions of agents' values and not the values
themselves. Therefore, this more general mechanism (that we call an
OPM with a restricted feasibility constraint) retains all of the good
properties of OPMs.

OPMs in multi-dimensional settings are similar: agents are approached
in turn (in an arbitrary order); each agent $i$ gets a price-menu over
the subset of services in $J_i$ that can be feasibly allocated to the
agent. However, we define SPMs differently: agents are approached in
turn (according to an optimal ordering) and offered individual
items at a time. Offers to a single agent are not necessarily
contiguous. These mechanisms are not truthful, but we show in
Section~\ref{subsec:undominated} that they can nonetheless be useful in
approximating the BMUMD.

\subsection{Myerson's optimal mechanism}
Myerson's seminal work describes the revenue maximizing mechanism for
the Bayesian single-parameter mechanism design problem. When the value
distributions $\disti$ are regular, in Myerson's mechanism the seller
first computes so-called virtual values for each agent, and then
allocates to a feasible subset of agents that maximizes the ``virtual
surplus''---the sum of the virtual values of agents in the set minus
the cost of serving that set of agents. We define these quantities
formally in Appendix~\ref{app:myerson}. For our analyses, we
mainly require the following two characterizations of the expected
revenue of any truthful mechanism when the value distributions are
regular. Similar characterizations hold in the non-regular case. These
and their extensions to the non-regular case are proved in
Appendix~\ref{app:myerson}.

\newcounter{myeCount}
\setcounter{myeCount}{\value{theorem}}
\begin{proposition}\label{lemma:Myerson}
When all input distributions $\disti$ are regular, the expected
revenue of any truthful single-parameter mechanism $M$ is equal to its
expected virtual surplus.
\end{proposition}

\newcounter{myeVsSumPqCount}
\setcounter{myeVsSumPqCount}{\value{theorem}}
\begin{lemma}
\label{lemma:revMyeVsSumPQ_regular}
If $\disti$ is regular for each $i$, for any truthful mechanism $M$
over the $n$ agents, the revenue of $M$ is bounded from above by
$\sum_i \pricemi \probmi$ where $\probmi$ is the probability (over
$\vali[1],\cdots,\vali[n]$) with which $M$ allocates to agent $i$ and
$\pricemi=\disti^{-1}(1-\probmi)$.

Furthermore for every $i$ (with a regular or non-regular value
distribution), there exist two prices $\underline{\pricei}$ and
$\overline{\pricei}$ with corresponding probabilities
$\underline{\probi}$ and $\overline{\probi}$, and a number $x_i\le 1$,
such that $x_i\underline{\probi} + (1-x_i)\overline{\probi} =
\probmi$, and the expected revenue of $M$ is no more than $\sum_i
x_i\underline{\pricei}\underline{\probi} +
(1-x_i)\overline{\pricei}\overline{\probi}$.
\end{lemma}


%% file: mpmd.tex
\label{subsec:md-redn}

We now present a general reduction from the multi-parameter optimal
mechanism design problem to the single-parameter
setting. Understanding the properties of optimal mechanisms in
multi-parameter settings is tricky. Our approach begins with an upper
bound on the optimal revenue in terms of the optimal revenue for a
related single-parameter problem following an approach of
\cite{CHK-07}. We describe this first.

\paragraph{An upper bound via copies.}
Consider an instance $\I$ of the BMUMD with $n$ agents, a set $J$ of
available services (with group $J_i$ of services targeted at agent
$i$), and a feasibility constraint $\sets$. We will define a new
instance of the BSMD in the following manner. We split each agent in
$\I$ into $|J_i|$ distinct agents (called ``copies''). Each copy is
interested in a single item $j\in J_i$ and behaves independently of
(and potentially to the detriment of) other copies. We call this
instance $\Ifc$. Formally, the instance has $|J|$ distinct agents
interested in a single service; agent $j$'s value for getting served,
$\vali[j]$, is distributed independently according to $\disti[j]$. The
mechanism again faces a feasibility constraint given by the set system
$\sets$.

$\Ifc$ is similar to $\I$ except that it involves more competition
(among different copies of the same multi-parameter agent). Therefore
it is natural to expect that a seller can obtain more revenue in the
instance $\Ifc$ than in $\I$. The following lemma formalizes this (see
Appendix~\ref{app:mp-redn} for a proof).

\begin{lemma}
\label{lem:mpmd-to-spmd}
Let $\Alg$ be any individually rational and truthful deterministic
mechanism for instance $\I$ of the BMUMD. Then the expected revenue of
$\Alg$, $\RevAlg$, is no more than the expected revenue of Myerson's
mechanism for the single-parameter instance with copies, $\Ifc$.
\end{lemma}

\paragraph{A reduction to single-dimensional OPMs.}
Next we show that if we can construct a good OPM for the setting with
copies, we can construct a good OPM for the multi-dimensional setting
as well. (Again, see Appendix~\ref{app:mp-redn} for a proof).

\begin{theorem}\label{theorem:opmBlanket}
Given an instance $\I$ of the BMUMD specified by the set system
$(J,\sets)$, there exists a truthful posted price mechanism for $\I$
which achieves an $\alpha$-approximation to the revenue achievable by
an optimal deterministic truthful mechanism, whenever there exists an
OPM for the corresponding BSMD instance $\Ifc$ that achieves an
$\alpha$-approximation to the optimal revenue for $\Ifc$.
\end{theorem}


%% file: matroid.tex
\label{sec:spm}

In this section we focus on the BSMD and present approximations to
optimal revenue via sequential posted price mechanisms for several
kinds of feasibility constraints, most notably matroids and matroid
intersections. Our exposition focuses on describing and analysing the
approximately-optimal SPMs, and we defer a discussion of efficiently
computing the SPMs to Appendix~\ref{app:compute}. While our focus is on revenue,
our techniques extend to a large class of objective functions, namely
those that are linear in the valuations of the served agents and the
payment received by the mechanism (see Appendix~\ref{app:other-objectives}).

While our analysis of the approximation factor depends closely on the
feasibility constraint, we use the same approach for constructing the
SPM in each case. We describe this next.

Suppose first that all the distributions $\disti$ are regular and do
not contain any point masses. Let $\probi=\probmi$ denote the
probability that Myerson's mechanism serves agent $i$, and let
$\pricei = \disti^{-1}(1-\probi)$ for all $i$. The SPM sets a price of
$\pricei$ for agent $i$ and offers to serve the agents in decreasing
order of their prices. If offered service, agent $i$ accepts with
probability exactly $\probi$. If the distribution $\disti$ contains
point masses, we modify the mechanism so that agent $i$ is offered the
price $\pricei$ with probability $\probi/(1-\disti(\pricei))$, and
again has a probability exactly $\probi$ of accepting the offer. We
denote this mechanism by $\spm$. We note that by Lemma~\ref{lemma:revMyeVsSumPQ_regular}, the
revenue of Myerson's mechanism is at most $\sum_i \pricei\probi$, and
we will compare the revenue of $\spm$ to this upper bound. Finally,
let the rank of a subset $S$ of agents, $\rank(S)$, denote the size of
the largest feasible subset in $S$, that is, $\rank(S)=\max_{S'\subset
S, S'\in \sets} |S'|$. Then, by definition,
$\sum_{i{\in}S}\probi\le\sum_{i{\in}S}\probmi\le\rank(S)$.

When the distributions are not regular, we pick prices $\prices$
randomly as suggested by Lemma~\ref{lemma:revMyeVsSumPQ_regular}.  In
Appendix~\ref{app:non-regular} we sketch the modifications required to
the analysis to obtain the same approximation factors for this case as
in the regular case.  We now present analyses for the expected revenue
of $\spm$ when all the distributions are regular.

\subsection{A 2 approximation for matroids}
\label{subsection:gooditem}
\label{sec:spm-matroid}

We first consider the setting where the set system $([n],\sets)$
is a matroid. Precisely, it satisfies the following conditions:
\begin{enumerate}
\item ({\bf heredity}) For every $A \in \sets$, $B \subset A$
  implies $B \in \sets$.
\item ({\bf augmentation}) For every $A, B \in \sets$ with $|A| >
  |B|$, there exists $e \in A\setminus B$ such that $B \cup \{e\} \in
  \sets$.
\end{enumerate}
Sets in $\sets$ are called independent, and maximal independent sets
are called {\em bases}. A simple consequence of the above properties
is that all bases are equal in size. Therefore, the rank of a set
$S\subseteq [n]$, is equal to the size of any maximal independent
subset of $S$. The {\em span} of a set $S\subseteq [n]$, $\spanOf(S)$,
is the maximal set $T\supseteq S$ with $\rank(T)=\rank(S)$.

We now show that for matroid set systems the SPM described above
approximates the expected revenue of the optimal mechanism within a
factor of 2.

\begin{theorem}
\label{thm:matroidSpm2Approx}
Let $\I$ be an instance of the BSMD with a matroid feasibility
constraint. Then, the mechanism $\spm$ described above
$2$-approximates the revenue of Myerson's mechanism for $\I$.
\end{theorem}
\begin{proof}
We show that the mechanism $\spm$ obtains an expected revenue of at
least $\frac 12\sum_i\pricei\probi$. Note that if the mechanism
ignored the feasibility constraint, and offered the prices $\prices$
to all agents, serving any agent that accepted its offered price, then
its expected revenue would be exactly $\sum_i\pricei\probi$. So our
proof accounts for the total revenue lost due to agents ``blocked''
from getting an offer by previously served agents.


Formally, let $\soldset=\{i_{1} < i_{2} < \dots < i_{\ell}\}$ be the
set of agents served, and let $\soldset_j$ denote the first $j$
elements of $\soldset$.  Define the sets
$\blockedset_{j}= 
\spanOf(\soldset_{j}) \setminus \spanOf(\soldset_{j-1})$. 
Note that the sets $\blockedset_{j}$ partition the set of blocked
agents. Moreover, $\blockedset_{j}\subseteq\{i:i \ge i_{j}\}$,
since we condition on serving $\soldset$, and so,
$\pricei\le\pricei[i_j]$ for all $i\in \blockedset_{j}$.

Denote the price offered to agent
$i_{j}$ by $\soldprice{j}$. Then, the expected revenue lost given that $\soldset$ is served is
\begin{align*}
\sum_{1\le j\le\ell}\sum_{i\in\blockedset_{j}}\pricei\probi
&\le\soldprice{1}\left(\sum_{i\in\spanOf(\soldset_1)}\probi\right)
+\sum_{1<j\le\ell}\soldprice{j}
\left(\sum_{i\in\spanOf(\soldset_{j})}\probi-\sum_{i\in\spanOf(\soldset_{j-1})}\probi\right)\\
&=\sum_{1\le j<\ell}\left((\soldprice{j}-\soldprice{j+1})\sum_{i\in\spanOf(\soldset_{j})}\probi\right)
+\soldprice{\ell}\left(\sum_{i\in\spanOf(\soldset_{\ell})}\probi\right)\\
&\le\sum_{1\le j<\ell}(\soldprice{j}-\soldprice{j+1})\cdot j + \soldprice{\ell}\cdot\ell
= \sum_{1\le j<\ell}\soldprice{j},
\end{align*}
which is the revenue obtained by serving $\soldset$. Here we used
$\sum_{i\in\spanOf(\soldset_j)}\probi\le\rank(\soldset_j)\le|\soldset_j|=j$. Therefore,
\begin{equation*}
\expect[\text{revenue lost}]
\le \sum_{\soldset}\sum_{j\in\soldset}\soldprice{j}\cdot\Pr[\text{$\soldset$ is served}]\\
=\RevSeq,
\end{equation*}
and so it follows that $\sum_{i}\pricei\probi\le2\RevSeq$.
\end{proof}

Blumrosen et al.~\cite{BH-08} show that the gap between the optimal SPM
and Myerson's mechanism can be as large as $\sqrt{\pi/2}\approx 1.253$
even in the single item auction case with \iid agents. We describe
this gap example in Appendix~\ref{app:spm-bad-ex}.

\subsection{Constant factor approximations for other feasibility constraints}
\label{sec:spm-other}

We now present improved approximations for special matroids, as well
as constant factor approximations for special non-matroid feasibility
constraints. The theorems below are proved in Appendix~\ref{app:improvedApprox}.

\paragraph{Uniform matroids and partition matroids.} A matroid is
$k$-uniform if all subsets of size at most $k$ are feasible. An
example of a $k$-uniform matroid constraint is a multi-unit auction
where the seller has $k$ units of an item on sale. We show that we can
obtain an improved $e/(e-1)\approx 1.58$ approximation in this
case. We show in Appendix~\ref{sec:spm-uniform} that this analysis
is tight. This result extends also to partition matroids,
i.e. disjoint unions of uniform matroids.

\begin{theorem}
\label{thm:uniform-spm}
Let $\I$ be an instance of the BSMD with a partition-matroid
feasibility constraint. Then, the mechanism $\spm$ described above
$e/(e-1)$-approximates the revenue of Myerson's mechanism for $\I$.
\end{theorem}

\input{spmMatInt.tex}

\input{non-matroid.tex}



%% file: spmMatInt.tex

\paragraph{Matroid intersections.} An intersection of $m$ matroids, 
$\mathcal{M}_1, \cdots, \mathcal{M}_m$, is a set system where a set is
feasible if and only if it is feasible in each of the $m$ matroids. An
example of an intersection of two matroids is a matching. We show that
the mechanism described above is an $m+1$ approximation for
intersections of $m$ matroids.

\begin{theorem}
\label{thm:intersection-spm}
Let $\I$ be an instance of the BSMD with a feasibility constraints
that is an intersection of $m$ matroids. Then, the mechanism $\spm$
described above $(m+1)$-approximates the revenue of Myerson's mechanism
for $\I$.
\end{theorem}

\paragraph{Combinatorial auctions with small bundles.} Consider a
situation where the seller has multiple copies of a number of items on
sale, and each agent is interested in some (commonly known) bundles
over items (and has a common value for all of these bundles). When
each desired bundle is of size at most $m$, we call this setting a
single-parameter combinatorial auction with known bundles of size
$m$. In this case the SPM described above achieves an $m+1$
approximation.

\begin{theorem}
\label{thm:bundles-spm}
Let $\I$ be an instance of a single-parameter combinatorial auction
with known bundles of size $m$. Then, the mechanism $\spm$ described
above $(m+1)$-approximates the revenue of Myerson's mechanism for
$\I$.
\end{theorem}

%% file: non-matroid.tex

\paragraph{The general non-matroid case.} We show in
Appendix~\ref{app:spm-nonmatroid} that the approximations described
above cannot extend to general non-matroid set systems. In particular,
the example we construct describes a family of instances with \iid
agents and a symmetric non-matroid constraint for which the ratio
between the expected revenue of Myerson's mechanism and that of the
optimal SPM is $\Omega(\log n/\log \log n)$ where $n$ is the number of
agents. The same example also shows that while in many
single-parameter pricing problems when the values are distributed in
the range $[1,h]$ it is possible to obtain an $O(\log h)$
approximation to social welfare, the same does not hold in our general
setting, and the gap can be $\Omega(h)$. On the other hand, the gap is
always bounded by $O(h)$ and is achieved by an SPM that charges each
agent a uniform price of $1$.

%% file: opm.tex
\label{sec:opm}
The approximations designed in Section~\ref{sec:spm} rely heavily on a
specific ordering of the agents. A natural question is whether the
seller can obtain good revenue when he has no control over the
ordering. In such a case the seller picks a set of prices in advance,
and then offers them to the agents on a first-come first-served
basis. We show that in many setting it is possible to determine a set
of prices for which such ``order-oblivious'' mechanisms (OPMs)
perform well.

As described in Section~\ref{sec:prelim}, an OPM specifies the prices
to charge every agent, as well as a feasibility constraint
(potentially different from $\sets$) to determine whether or not to
make an offer to an agent. To pick the prices, we follow the approach
taken in Section~\ref{sec:spm}. The prices in the OPM are either set
to be the same as for the corresponding approximately-optimal SPM, or
set to infinity (effectively dropping the respective agent from
consideration). We now present the details for different kinds of
feasibility constraints.

\subsection{An $O(\log k)$ approximation for general matroids}
\label{subsection:opmMatroid}
For general matroids we give an $O(\log k)$ bound on the gap below,
where $k$ is the rank of the matroid. We remark that a similar result
was obtained by Babaioff et al.~\cite{BIK-07} for the related matroid
secretary problem. However, we show in
Appendix~\ref{app:opm-general-matroid} that their approach cannot give
a non-trivial approximation in our setting.

\begin{theorem}
\label{lemma:kApproxMatroid}Let $\I$ be an instance of the BSMD with a matroid feasibility
constraint. Then, there exists a
set of prices $\prices$ such that $\RevObl$ $O(\log k)$-approximates $\RevMye$ for $\I$.
\end{theorem}
\begin{proof}
We present the proof for regular distributions. Appendix~\ref{app:non-regular} presents the extension to the non-regular case.
Note that since the feasibility constraint is a matroid, for any
instantiation of values, the worst (least revenue) allocation is
achieved when agents arrive in the order of increasing
prices. Hereafter we assume that agents always arrive in that order.
Let $\offerProb$ be as defined in
Section~\ref{subsection:postedPriceMech}; recall that the expected
revenue may be expressed as $\sum_{i}\offerProb\pricei\probi$.  

Now consider a hypothetical situation where the prices are all equal
to $1$ but the probabilities with which the agents accept the offered
prices are still $\probi$. Then, the expected revenue of this
hypothetical mechanism would be given by $\sum_i c_i\probi$ which is
at least $1/2 \sum_i\probi$ by the argument in
Theorem~\ref{thm:matroidSpm2Approx}. In other words, the weighted
average of the $c_i$s is at least $1/2$, weighted by the $\probi$s. We
get the following sequence of implications.
\begin{align*}
(1/2)\sum_{i}\probi \le \sum_i\offerProb\probi \le \sum_{i:\offerProb<1/4} \probi/4
+ \sum_{i:\offerProb\ge 1/4} \probi = (1/4)\sum_{i}\probi +
(3/4)\sum_{i:\offerProb\ge 1/4}\probi \
\Rightarrow \ \sum_{i:\offerProb \geq 1/4}\probi \geq (1/3)\sum_{i}\probi
\end{align*}
This means that the probability mass of elements having $\offerProb
\geq 1/4$ is at least a third of the total. Let 
    $\goodSet = \{i \vert \offerProb \ge 1/4\}$;
the revenue obtained from serving only the agents in $\goodSet$ is
\begin{align}
\label{eq:revOfGoodSet}
    \sum_{i\in \goodSet}\offerProb\pricei\probi
    &\ge 1/4\sum_{i \in \goodSet}\pricemi\probmi.
\end{align}

Consider recursively applying the above argument to the elements
outside $\goodSet$.  At step $j$, let $\goodSet_j$ be the newly found
$\goodSet$, 
and let $\matroidElts_j$ be the
set of agents still under consideration, defined as
   $\matroidElts_1=[n]$  and 
   $\matroidElts_j=\matroidElts_{j-1}-\goodSet_{j-1}$ for $j>1$.
Now, at each stage, $\goodSet_j$ contains at least one third of the
total probability mass of the remaining elements; thus, at stage
$\ell=\lceil 1+\log_{3/2} k\rceil,$ we would have reduced the total probability
mass to less than $3/4$; by noting that any singleton
set is independent in a matroid and applying Markov's inequality we
may see that $\goodSet_\ell=\matroidElts_\ell$.  Since the collection of
$\goodSet_j$'s form a size $O(\log k)$ partition of $[n]$, and
summing~\eqref{eq:revOfGoodSet} over the collection gives a total
expected revenue of $\RevMye/4$,  we may conclude that there is some
$\goodSet_j$ which gives a $\Omega(1/\log k)$-fraction of $\RevMye$
regardless of ordering.
\end{proof}

We remark that while the $2$-approximate SPM in Section~\ref{sec:spm}
can be computed efficiently, we do not know of an efficient algorithm
for computing an $O(\log k)$-approximate order-oblivious pricing.

\subsection{Improved approximations for special matroids}
\label{subsection:opmSpecialMatroid}

We first note that for the case of uniform matroids (where every set
of size at most $k$ is independent), 
an approximation of $3$ can be obtained
by employing techniques developed by Chawla, Hartline and Kleinberg
\cite{CHK-07} for pricing problems in multi-parameter settings.  We
can further improve this approximation factor to $2$ via techniques
developed by Samuel-Cahn \cite{Cahn84} in the context of prophet
inequalities in optimal stopping theory. We describe this approach in 
Appendix~\ref{app:opm-uniform} and show that this approximation
factor is tight.


\begin{theorem}
  \label{thm:uniformOpmApprox} Let $\I$ be an instance of the BSMD
with a uniform matroid feasibility constraint. Then, there exists a
set of prices $\prices$ such that $\RevObl$ $2$-approximates
$\RevMye$ for $\I$.
\end{theorem}

\begin{corollary}
Let $\I$ be an instance of the BSMD with a partition matroid
feasibility constraint. Then, there exists a set of prices $\prices$
such that $\RevObl$ $2$-approximates $\RevMye$ for $\I$.
\end{corollary}





For graphical matroids, Babaioff et al. \cite{BDGIT09} and Korula and
P\'{a}l \cite{KP09} develop approaches for reducing this case to a
partition matroid that in our setting yield a $4$-approximation to the
optimal revenue; in Appendix~\ref{app:graphicalMatroid} we use a
similar approach but exploit the connection between prophet
inequalities and partition matroids to obtain a $3$-approximation.

\begin{theorem}
\label{thm:graphicalOpmApprox}
Let $\I$ be an instance of the BSMD with a graphical matroid
feasibility constraint. Then, there exists an OPM with a restricted
feasibility constraint that is a partition matroid, that
$3$-approximates $\RevMye$ for $\I$.
\end{theorem}

\subsection{OPMs for matroid intersections}
\label{subsection:opmMatroidIntersection}
As for SPMs, our techniques for approximately-optimal OPMs in matroid
settings extend to intersections of few matroids. For intersections of
two partition matroids we get an $6.75$-approximation (see theorem below,
and proof in Appendix~\ref{app:opm-intersection}). For intersections
of $m$ arbitrary matroids, our techniques imply an $O(m\log k)$
approximation where $k$ is the maximum size of a feasible set (and so
is bounded by the least matroid rank); we omit the proof for brevity.
\begin{theorem}\label{thm:opmPartitionMatroidIntersection}
Let $\I$ be an instance of the BSMD with a feasibility constraint
given by the intersection of two partition matroids. Then, there
exists a set of prices $\prices$ such that $\RevObl$ $6.75$-approximates
$\RevMye$ for $\I$.
\end{theorem}

\paragraph{The non-matroid case.}
Example~\ref{example:spm-non-matroid} in
Appendix~\ref{app:spm-nonmatroid} already implies that
order-oblivious pricings cannot obtain more than an $O(\log
n/\log\log n)$ fraction of the revenue of Myerson's mechanism in
general in non-matroid settings. How do they compare to the optimal
SPM? We show in Appendix~\ref{app:opm-non-matroid} that the gap
between the optimal order-oblivious pricing and the optimal SPM can
be large --- $\Omega(\log n/\log\log n)$ --- in the non-matroid
setting.

%% file: mp-approx.tex
\label{sec:mp-approx}
We now present approximations for various versions of the BMUMD.

\subsection{Approximation through truthful mechanisms}

We first note that for the hotel rooms example discussed in the
introduction, and indeed for any setting with unit-demand agents and
multiple units of multiple items on sale, a $6.75$-approximation
follows from Theorems~\ref{theorem:opmBlanket}
and \ref{thm:opmPartitionMatroidIntersection}.

\begin{theorem}
Consider an instance of the BMUMD where the seller has multiple copies
of $n$ items on sale, and agents are unit-demand and have
independently distributed values for each item. Then there exists an
$6.75$-approximate OPM for this instance. The prices for this mechanism
can be computed in polynomial time.
\end{theorem}

\noindent
A similar result for graphical matroids follows from
Theorems~\ref{theorem:opmBlanket} ,~\ref{thm:graphicalOpmApprox}
and \ref{thm:opmPartitionMatroidIntersection} (see Appendix~\ref{app:sec6-proofs} for
a proof).

\begin{theorem}
\label{thm:bmumd-graphical}
Consider an instance of the BMUMD based on a graph $G=(V,E)$ where the
agents have independent values for different edges and are interested
in buying one edge each. The seller can allocate any forest in the
graph. Then there exists a 10.67 approximate OPM for this instance. The
prices for this mechanism can be computed in polynomial time.
\end{theorem}

\subsection{Approximation through implementation in undominated strategies}
\label{subsec:undominated}

Both of the results above involve feasibility constraints that admit
good OPMs in single-dimensional settings. Can we design good
multi-dimensional mechanisms for set systems that admit good SPMs in
the single-dimensional setting, but for which we do not know constant
approximate OPMs? Two examples are general matroid intersections and
combinatorial auctions with small bundles (e.g. the airline tickets
setting described in Section~\ref{sec:prelim}).

We now show that this can be done if we relax truthfulness to {\em
implementation in undominated strategies}. (See formal definition in
Appendix~\ref{app:sec6-proofs}.)  Our mechanism for both of the cases
above is an SPM specified by a set of prices, one for each service in
$J$, and an ordering over services. It begins by announcing the prices
to the agents. Then, as for single-dimensional instances, it considers
offering the services to the agents in turn: at every step, depending
on the services allocated so far, it determines whether or not it is
feasible to allocate the next service $j\in J_i$ to the corresponding
agent $i$, and if so, offers a price $\pricei[j]$ to $i$. This
mechanism is not truthful. For example, an agent may reject an offer
for a service $j$ even if his value for $j$ exceeds its price, if he
anticipates obtaining a more profitable offer in the
future. Nevertheless we can infer some properties about rational agent
behavior in such a mechanism.

\begin{lemma}
\label{lem:undominated}
Consider an instance $I$ of the BMUMD and an SPM as defined above with
prices $\prices$ and ordering $\sigma$. Then, the following holds for
any undominated strategy of any agent: if an agent $i$ desires only
one service at prices $\prices$, that is, $\vali[j]\ge \pricei[j]$ for
only one $j\in J_i$, then the agent must accept $j$ if offered the
service.
\end{lemma}

\noindent
This lemma and the following theorem are proved in Appendix~\ref{app:sec6-proofs}.

\begin{theorem}
\label{thm:mdmd-spm}
Given an instance of the BMUMD with a general matroid intersection
constraint, there exists an SPM that implements a $8$-approximation in
undominated strategies. Given an instance of a combinatorial auction
with known bundles of size $2$, there exists an SPM that implements a
$8$-approximation for the instance in undominated strategies.
\end{theorem}

%% file: discuss.tex
We presented constant factor approximations to revenue for several
classes of multi-dimensional mechanism design problems by designing
approximately-optimal posted price mechanisms for single-dimensional
settings. This approach does not extend beyond matroid and
matroid-like settings. However, it is possible that there is some
other class of simple near-optimal mechanisms for non-matroid
single-dimensional settings that do not exploit competition among
agents. Such mechanisms may lead to approximately-optimal
multi-dimensional mechanisms for a broader class of feasibility
constraints. 

More generally, two important assumptions underlie our work: (1)
agents are unit-demand, and (2) their values for different services
are distributed independently. In the absence of either of these
assumptions the upper bound on the optimal revenue based on the
setting with copies does not remain valid. An important open question
is to design a reasonably tight upper bound in those cases, and use
it to approximate the optimal mechanism.

%% file: appendix.tex
\input{myerson-appendix.tex}

\section{Reducing BMUMD to BSMD}
\label{app:mp-redn}

In this section we prove Lemma~\ref{lem:mpmd-to-spmd} and Theorem~\ref{theorem:opmBlanket}.

\vspace{0.2in}
\begin{proofof}{Lemma~\ref{lem:mpmd-to-spmd}}
We first note that a mechanism is individually rational if we have
$\pi_i\le \vali[j]$ for $j\in S\cap J_i$, and $\pi_i=0$ if $S\cap
J_i=\emptyset$. Truthful mechanisms in multi-parameter settings
satisfy the weak monotonicity condition defined below.

\begin{definition}
\label{def:weak-mon}
A mechanism $(M,\pi)$ satisfies weak monotonicity if for any agent $i$
and any two types (value vectors) $\vals^1$ and $\vals^2$ with
$\val^1_j = \val^2_j$ for all $j\in J\setminus J_i$, the following
holds:
\[ \val^1_{M_i(\vals^1)}+ \val^2_{M_i(\vals^2)} \ge
\val^1_{M_i(\vals^2)}+ \val^2_{M_i(\vals^1)} \]
Here $M_i(\vals)$ denotes the unique index in $M(\vals)\cap J_i$.
\end{definition}

We show that we can construct a truthful mechanism $\Algfc$ for the
$\Ifc$ with revenue no less than that of $\Alg$. The lemma then
follows from the optimality of Myerson's mechanism. Given a vector of
values $\vals$, the mechanism $\Algfc$ allocates to the set that
$\Alg$ allocates to in $\I$ given the same value vector. We first
claim that the allocation rule of $\Algfc$ is monotone non-decreasing
in any $\vali[j]$, implying that there exists a payment rule that
makes the mechanism truthful. To prove the claim, fix any agent $i$
and $j\in J_i$, and consider two value vectors $\vals^1$ and $\vals^2$
with $\val^1_{j}=x$, $\val^2_{j}=y$, and $\val^1_{j'} = \val^2_{j'}$
for $j'\ne j$. Let $\alpha_x$ and $\alpha_y$ denote the probabilities
of serving agent $i$ with service $j$ under the two value vectors
respectively, and let $\beta_x$ and $\beta_y$ denote the total value
that agent $i$ obtains from other services $j'\in J_i$, $j'\ne j$, in
the two cases respectively. Then the weak-monotonicity
(Definition~\ref{def:weak-mon}) of $\Alg$ implies that
\[(x\alpha_x+\beta_x) + (y\alpha_y+\beta_y) \ge (x\alpha_y+\beta_y) +
(y\alpha_x+\beta_x)\]
or,
\[(x-y)(\alpha_x-\alpha_y)\ge 0\]
Therefore the claim holds.

It remains to prove that the expected revenue of $\Algfc$ given $\Ifc$
is no less than the expected revenue of $\Alg$ given $\I$. Note that
any deterministic multi-parameter mechanism can be interpreted as
offering a price menu with one price for each item or service to each
agent as a function of other agents' bids \cite{Wilson-book}. The
agent then chooses the item or service that brings her the most
utility. Given this characterization, suppose that for a fixed set
$\vals$ of values, mechanism $\Alg$ offers a price menu with prices
$\rprices$ to agent $i$. Then, it draws a revenue of $\rpricei[j]$
from $i$ whenever service $j$ is offered. On the other hand, mechanism
$\Algfc$ charges the agent $j$ the minimum amount it needs to bid to
be served, which is no less than $\rpricei[j]$, as $\Alg$ is
individually rational.
\end{proofof}

\vspace{0.2in}

\begin{proofof}{Theorem~\ref{theorem:opmBlanket}}
Consider an $\alpha$-approximate OPM for $\Ifc$ with prices
$\prices$. The $\alpha$-approximate mechanism for $\I$ is described as
follows. It serves the agents in the order in which they arrive. When
agent $i$ arrives, depending on the set of services already allocated,
the mechanism determines the subset $J'_i$ of services in $J_i$ that
can be feasibly allocated to $i$, and offers a price menu of
$\{\pricei[j]\}_{j\in J'_i}$ to $i$. Agent $i$ then chooses a service from
the menu and this service is allocated to it. Truthfulness follows
from the definition. In order to argue that the mechanism is
$\alpha$-approximate, we will show that its revenue is no less than
the revenue of the OPM for $\I$---$\RevObl$. Then the result follows
from Lemma~\ref{lem:mpmd-to-spmd}. To see that the expected revenue of
the mechanism is at least $\RevObl$, we claim that the mechanism
allocates a maximal feasible set of services. If not, then there
exists an agent $i$ and a service $j$ such that it is feasible to
allocate $j$ to $i$ (that is, $j\in J'_i$, and $i$ has not be
allocated any service), and the value of $i$ for $j$ exceeds its
price. Then, at the time that $i$ is offered a price menu, it must
have been the case that $i$ chose $j$ or some other service in $J'_i$
and got allocated that service, and we get a contradiction. This
concludes the proof.
\end{proofof}

\section{Approximations via SPMs}
\label{app:improvedApprox}
In this section we present missing proofs from
Section~\ref{sec:spm}. In particular, we prove that the SPMs described
in Section~\ref{sec:spm} are $e/(e-1)$ approximate for partition
matroids, $m$ approximate for intersections of $m$ matroids, and $m$
approximate for combinatorial auctions with known bundles of size $m$.

\subsection{A lower bound example for $1$-uniform matroids}
\label{app:spm-bad-ex}
\input{spm-lb.tex}

\subsection{Proof of Theorem~\ref{thm:uniform-spm}: an $\frac{e}{e-1}$ approximation for uniform and partition matroids}
\label{sec:spm-uniform}

\input{uniform-approx.tex}

\subsection{Proof of Theorem~\ref{thm:intersection-spm}: an $m+1$
  approximation for intersections of $m$ matroids}
\label{app:spm-intersection}


Let the $m$ matroids be denoted by
$\mathcal{M}_1,\ \mathcal{M}_2,\dots,\ \mathcal{M}_m$. Let
$\rank_a(S)$ and $\spanOf_a(S)$ denote respectively the rank and span
of set $S$ in the matroid $\mathcal{M}_a$. Note that for any subset
$S$ and any $a\in [m]$, we have $\sum_{i{\in}S}\probmi\le\rank_a(S)$.



Once again, let $\soldset=\{i_{1} < i_{2} < \dots < i_{\ell}\}$ denote
the set of agents served.  We prove the theorem by showing that the
expected revenue of $\spm$ is at least $1/(m+1)\sum_{i}\pricei\probi$,
by arguing that the total price paid by agents in $\soldset$ is at
least $1/m$ times the expected revenue from agents that are
``blocked'' by $\soldset$.

Let $\soldset_j$ denote the first $j$ elements of $\soldset$.
For each $1{\le}a{\le}m$, define sets $\blockedset_{j}^{a}$ with
respect to matroid $\mathcal{M}_a$ as in the proof of
Theorem~\ref{thm:matroidSpm2Approx}. That is, $\blockedset_j^a =
\spanOf_a(\soldset_j)\setminus\spanOf_a(\soldset_{j-1})$. Denote the
price of item $i_{j}$ by $\soldprice{j}$.  Then, if we let
$\blockedset_{j}=\cup_{a=1}^{m}\blockedset_{j}^{a}$, we can upper
bound the expected revenue lost when $\soldset$ is served by
\begin{equation*}
\sum_{1\le j\le\ell}\sum_{i\in\blockedset_{j}}\pricei\probi
\le\sum_{a=1}^{m}\sum_{1\le j\le\ell}\sum_{i\in\blockedset_{j}^{a}}\pricei\probi
\le m\sum_{1\le j<\ell}\soldprice{j}.
\end{equation*}
Here we used the same algebraic transformation as in the proof of
Theorem~\ref{thm:matroidSpm2Approx} along with the fact that
$\sum_{i\in\blockedset_j^a}\probi\le\sum_{i\in\spanOf_a(\soldset_j)}\probi\le
j$.

Therefore as before we get 
$\sum_{i}\pricei\probi\le(m+1)\RevSeq$.

\subsection{Proof of Theorem~\ref{thm:bundles-spm}: an $m+1$
  approximation for combinatorial auctions with known bundles of size $m$}

Let $A$ denote the set of items available to the seller, each with
some multiplicity. First suppose that each agent is single-minded,
that is, each agent is interested in only one bundle of items, the
bundle being of size at most $m$. Then, the feasibility constraint is
an intersection over $|A|$ uniform matroids, one corresponding to each
item, with each agent participating in only $m$ of the matroids. Now
it is easy to adapt the proof of Theorem~\ref{thm:intersection-spm} to
obtain an $m+1$ approximation.

More generally suppose that every agent is interested in a collection
of bundles, each of size at most $m$, and modify the mechanism $\spm$
so that in addition to deciding whether or not to serve an agent, it
also arbitrarily allocates any available desired bundle to every agent
it serves. Then we can argue that for any set $S$, and set $B$ blocked
by the agents in $S$, the sum of the probabilities $\probi$ over the
set $B$ is no more than $m$ times the size of $S$. Therefore, once
again following along the proof of Theorem~\ref{thm:intersection-spm},
we get an $m+1$ approximation.

\subsection{Bad gap example for general non-matroids}
\label{app:spm-nonmatroid}

We now show that the approximations described above cannot extend to
general non-matroid set systems. In particular, the example below
describes a family of instances with \iid agents and a symmetric
non-matroid constraint for which the ratio between the expected
revenue of Myerson's mechanism and that of the optimal SPM is
$\Omega(\log n/\log \log n)$ where $n$ is the number of agents.

\begin{example}
\label{example:spm-non-matroid}
For a given $m$ set $n=m^{m+1}$. Partition $[n]$ into $m^m$ groups
$G_1, \cdots, G_{m^m}$ of size $m$ each, with $G_i\cap G_j=\emptyset$
for all $i\ne j$. The set system $\sets$ contains all subsets of
groups $G_i$, that is, $\sets = \{ A: \exists i \text{ with }
A\subseteq G_i \}$. Each agent has a value of $1$ with probability
$1-1/m$ and $m$ with probability $1/m$.  

For any given valuation profile, let us call the agents with a value
of $m$ to be good agents and the rest to be bad agents. The
probability that a group contains $m$ good agents is
$m^{-m}$. Therefore in expectation one group has $m$ good agents and
Myerson's mechanism can obtain revenue $m^2$ from such a group:
$\RevMye=\Omega(m^2)$.

Next consider any SPM. The mechanism can serve at most $m$ agents. If
all the served agents are bad, the mechanism obtains a revenue of at
most $m$. On the other hand, once the mechanism commits to serving a
good agent, it can only serve agents within the same group in the
future. These have a total expected value less than $2m$. Therefore,
the revenue of any SPM is at most $3m$, and we get a gap of $\Omega(m)
= \Omega(\log n/\log\log n)$.
\end{example}

The above example also shows that while in many single-parameter
pricing problems when the values are distributed in the range $[1,h]$
it is possible to obtain a $\log h$ approximation to social welfare,
the same does not hold in our general setting. In the example we have
$h=m$ and the gap between the expected revenue of the optimal SPM and
that of Myerson's mechanism is $\Omega(h)$. On the other hand, the gap
is always bounded by $O(h)$ and is achieved by an SPM that charges
each agent a uniform price of $1$.

\section{Approximations via OPMs}

\subsection{General matroids.}
\label{app:opm-general-matroid}
In Section~\ref{subsection:opmMatroid} we design an $O(\log k)$
approximate OPM for general matroids.  We remark that a similar result
was obtained by Babaioff et al.~\cite{BIK-07} for the related matroid
secretary problem. In Babaioff et al.'s setting agents arrive in a
random order but their values are adversarial. They present an $O(\log
k)$ approximation by picking a price uniformly at random in the set
$\{h/k, 2h/k, \cdots, h\}$ and charging it to every agent; here $h$ is
the largest among all values. In our setting such an approach does not
work: the example below shows that no uniform pricing can achieve an
$o(\log h)$ approximation even for $k=1$.

\begin{example}
Let $k=1$ and consider a group of $h$ agents where agent $i$ has a
value of $i$ with probability $1/2i^2$ and zero otherwise. Then an SPM
that sets a price of $i$ for agent $i$ obtains an expected revenue of
$\Omega(\log h)$. On the other hand, an SPM that uses a uniform price
of $c$ only obtains expected revenue $\sum_{i\in [c,h]} c/2i^2<
c/2c=1/2$.
\end{example}

\subsection{Uniform and partition matroids}
\label{app:opm-uniform}

\input{prophet.tex}

\input{uniform-opm.tex}

\input{opm-lb.tex}

\subsection{Graphical matroids}
\label{app:graphicalMatroid}

\input{graphical-opm.tex}

\subsection{Matroid intersections}
\label{app:opm-intersection}

\input{matroidInt-opm.tex}

\subsection{Order-oblivious pricings in the non-matroid setting}
\label{app:opm-non-matroid}

In this section we present an example with a non-matroid constraint
for which the revenue obtained by ordering the agents in the optimal
way is a factor of $\Omega(\log n/\log\log n)$ larger than that
obtained by ordering the agents in the least optimal way.

\begin{lemma}
There exists an instance of the single-parameter mechanism design
problem with a non-matroid feasibility constraint, along with two
orderings $\sigma_1$ and $\sigma_2$ such that the revenue of the
optimal SPM using ordering $\sigma_1$ is a factor of $\Omega(\log
n/\log\log n)$ larger than that of the optimal SPM using ordering
$\sigma_2$.
\end{lemma}

\begin{proof}
Consider the following example.  Construct a complete $m$-ary tree of
height $m+1$, and place a single agent at each node other than the
root.  The agents' valuations are \iid, where any agent has a
valuation of $m$ with probability $1/m$, and a valuation of $0$
otherwise.  Our constraint on serving the agents is that we may serve
any set of agents that lie along a single path from the root of the
tree to some leaf -- it is easy to verify that this is
downward-closed.

Consider what happens when we may serve the agents in order by level
from the root of the tree to the leaves.  At each level of the tree, we
may offer to serve at least $m$ different agents, regardless of the
outcome on previous levels.  Since we may never sell to more than one
agent per level, our revenue is either $0$ or $m$ on each level.  We
get a revenue of $0$ if and only if every agent has a valuation of
$0$; this occurs with probability at most
\begin{equation*}
(1-1/m)^m \le 1/e,
\end{equation*}
and thus our expected revenue overall is at least
\begin{equation*}
m^2\cdot(1-1/e) = \Omega(m^2).
\end{equation*}

On the other hand, if we must serve the agents in order by level from
the leaves of the tree to the root, then the first agent we serve
commits us to a specific path.  So we cannot hope to achieve revenue
better than $m$ for this specific node, plus the revenue expected
revenue for an arbitrarily chosen path.  Since each agent has an
expected valuation of $1$, this is bounded by
\begin{equation*}
m + (m-1)\cdot1 = O(m).
\end{equation*}

Thus, the difference in revenue between the described orderings is
$\Omega(m)$; since the total number of agents is $n=O(m^{m})$, in
terms of $n$ this gap is $\Omega(\log n/ \log\log n)$.
\end{proof}

\section{Approximation in the non-regular case}
\label{app:non-regular}

We now sketch changes required to the theorems proved in
Sections~\ref{sec:spm} and \ref{sec:opm} to obtain the same
approximations in the non-regular case. 

From Lemma~\ref{lemma:revMyeVsSumPQ_regular} we know that for every
$i$, there exist prices $\underline{\pricei}$ and $\overline{\pricei}$
with corresponding probabilities $\underline{\probi} =
1-\disti(\underline{\pricei})$ and $\overline{\probi} =
1-\disti(\overline{\pricei})$, as well as a number $x_i$ such that
$x_i\underline{\probi}+(1-x_i)\overline{\probi} = \probmi$, and
Myerson's expected revenue is bounded by $\sum_i
(x_i\underline{\pricei}\underline{\probi}
+(1-x_i)\overline{\pricei}\overline{\probi})$. In fact this holds more
generally. Let $\probi$ be any probability less than $1$. Then there
exist probabilities $\underline{\probi}$ and $\overline{\probi}$, and
a number $x_i\in [0,1]$ with $\probi =
x_i\underline{\probi}+(1-x_i)\overline{\probi}$, such that for
$\underline{\pricei}=\disti^{-1}(1-\underline{\probi})$,
$\overline{\pricei}=\disti^{-1}(1-\overline{\probi})$, and $\pricei$
defined as
\[\frac{x_i\underline{\pricei}\underline{\probi}
+(1-x_i)\overline{\pricei}\overline{\probi}}{\probi},\]
the optimal revenue achievable by selling an item with probability
$\probi$ to agent $i$ is no more than $\pricei\probi$.


Now consider a hypothetical situation in which the probability that
agent $i$ accepts a price of $\pricei$ is exactly $\probi$, and
consider running an SPM/OPM with prices $\pricei$ that is
$\alpha$-approximate with respect to $\sum_i \pricei\probi$. We claim
that we obtain an $\alpha$-approximation by using the same SPM/OPM but
instead picking a price of $\underline{\pricei}$ with probability
$x_i$ and $\overline{\pricei}$ with probability $1-x_i$.

To prove the claim we first note that we can defer the process of
picking a price for agent $i$ until the mechanism decides to offer the
agent some price. In this case, the probability that the agent accepts
the offered price is exactly $\probi$, and the revenue obtained from
the agent conditioned on serving him is exactly $\pricei$. Therefore,
the probability that the mechanism makes an offer to an agent is also
identical to the corresponding probability in the hypothetical
deterministic mechanism, and the expected revenue of the mechanism is
exactly the same as that of the hypothetical mechanism.

\section{Computing the near-optimal posted-price mechanisms}
\label{app:compute}

We now describe how to compute the approximately optimal OPMs and SPMs
designed in Sections~\ref{sec:spm} and \ref{sec:opm}. We assume that
we are given access to the following oracles and algorithms:
\begin{itemize}
\item An algorithm to compute the optimal price to charge to a
  single-parameter agent given the agent's value distribution. Note
  that given such an algorithm and some value $x$, we can modify it to
  return the optimal price in the range $[x,\infty)$ to charge the
    agent.
\item An oracle that given a value $v$ and index $i$ returns
  $\disti(v)$ and $\densi(v)$, as well as, given a probability
  $\alpha$ returns $\disti^{-1}(\alpha)$. Note that the oracle can be
  used to compute the virtual value $\virti(v)$.
\item An oracle for computing ironed virtual values in order to
  compute the approximately optimal SPM for non-regular distributions.
\item An algorithm to maximize social welfare over the given
  feasibility constraint in order to be able to compute the outcome of
  Myerson's mechanism.
\end{itemize}

All of the mechanisms designed by us require computing the
probabilities $\probmi$. We first show how to estimate these
probabilities within small constant factors:
\begin{enumerate}
\item Let $\epsilon=1/3n$. Sample $N=4n^4\log n/\epsilon^2$ value
profiles from $\disti[1] \times \disti[2] \times \cdots 
\times\disti[n]$. For each sample, compute the (ironed) virtual value
for each agent, and use these to compute the outcome of Myerson's
mechanism for that value profile.
\item Estimate the probabilities $\probmi$ using the samples. Call the
estimates $\widehat{\probmi}$.
\item If $\widehat{\probmi}<1/n^2$, set $\widehat{\probi}=1/n^2$, else set 
$\widehat{\probi} = \widehat{\probmi}/(1-\epsilon)$.
Compute for each $i$ the value
$\widehat{\pricei}=\disti^{-1}(1-\widehat{\probi})$.
\item Find the optimal price in the range $[\widehat{\pricei},\infty)$ to
charge to agent $i$. Call it $\pricei$. Let $\probi=1-\disti(\pricei)$.
\item Output the prices computed in the last step and order the agents
in order of decreasing prices.
\end{enumerate}

\noindent
In order to analyse the performance of this approach, we compare it to
a mechanism that charges agent $i$ the price $\pricemi
= \disti^{-1}(1-\probmi)$ but uses the same ordering as the mechanism
above.  We first show that the probabilities $\probi$ closely estimate
the probabilities $\probmi$.



\begin{lemma}
\label{lem:estimates}
With probability at least $1-2/n$, we have $\widehat{\probi}\in
[\probmi,(1+3\epsilon)\probmi+2/n^2]$.
\end{lemma}
\begin{proof}
First, for any $i$ with $\probmi\ge 1/n^4$, using Chernoff bounds we
get that
\begin{align*}
{\textbf{Pr}}\big[\big| \widehat{\probmi}-  \probmi\big| \geq \epsilon\probmi\big]  \leq
2e^{-\epsilon^2\probmi N/2} \le 2/n^2
\end{align*}
$\widehat{\probmi}\in(1\pm\epsilon)\probmi$ in turn implies by definition
that $\probmi\le \widehat{\probi}\le
(1+\epsilon)/(1-\epsilon)\probmi\le
(1+3\epsilon)\probmi$. Therefore we have $\widehat{\probi}
\in [\probmi,(1+3\epsilon)\probmi]$. On the other hand, for
$\probmi< 1/n^4$, by Markov's inequality, with probability $1-1/n^2$,
$\widehat{\probmi}<1/n^2$, and so $\widehat{\probi}\in [\probmi,1/n^2]$. The
lemma now follows by employing the union bound.
\end{proof}

Furthermore, conditioned on the event defined in the statement of the
above lemma (call it $\mathcal E$), since $\pricemi$ lies in the range
$[\widehat{\pricei},\infty)$, we have that
$\probmi\pricemi \le \probi\pricei$. This implies that the prices
$\pricei$ give a good estimate on the revenue of Myerson's mechanism.

Next, we compare the real mechanism $\spm$ with prices $\pricei$ to
the theoretically good mechanism $\spmp$ that charges prices
$\pricemi$. Let $S$ be the set of agents for which
$\widehat{\probmi}<1/n^2$. The probability that any of these agents is
offered service in $\spm$ is at most $1/n$. Conditioned on this event
not happening, the probability that an agent is made an offer in
$\spm$ is no smaller than its counterpart in $\spmp$. Moreover,
conditioned on being made an offer, the revenue from an agent $i$ is
$\probi\pricei\ge \probmi\pricemi$.

Therefore, conditioned on the event $\mathcal E$, the expected revenue
of $\spm$ is at least a $(1-1/n)$ fraction of the expected revenue of
$\spmp$. But the event $\mathcal E$ happens with probability $1-2/n$,
therefore, we get a $(1-o(1))$ approximation to the expected revenue
of $\spmp$.

\section{Approximations for the BMUMD}
\label{app:sec6-proofs}

We first prove that there exists a good OPM for instances of the BMUMD
involving a feasibility constraint that is the intersection of a
graphical matroid and the agents' unit-demand constraints.

\vspace{0.2in}
\begin{proofof}{Theorem~\ref{thm:bmumd-graphical}}
Note that though the feasibility constraint we are facing is the intersection
of a graphical matroid and partition matroid (from the unit
demand constraint), we can view the situation as if we were in the intersection
of two partition matroids. This follows from the proof of 
Theorem~\ref{thm:graphicalOpmApprox}, where we see that a graphical matroid can be 
seen as a union of 1-uniform matroids, which is a partition matroid. The total
probability mass of the elements of each 1-uniform matroid is at most 2. Thus, if we sell 
at prices for which the probability of an agent $i$ desiring the item is $\probmi/4$,
then with a probability of at least 1/2 no more than 1 agent will desire service in the 
1-uniform matroid which contains $i$ and with a probability of at least 3/4 no more than 1 item 
is desired by the agent $i$. Thus the revenue obtained gives an approximation factor of $4\cdot4/3\cdot2 = 32/3\approx10.67$.
\end{proofof}

\vspace{0.2in}

\noindent
Next we prove that for the two settings discussed in
Section~\ref{subsec:undominated}, we can design an SPM that achieves
a good approximation via implementation in undominated strategies.

Formally, for an agent $i$, a strategy $s_i$ is said to be dominated
by a strategy $s'_i$ if for all strategies $s_{-i}$ of other agents,
the utility that $i$ obtains from using $s_i$ is no better than that
from using $s'_i$, and for some strategy $s_{-i}$, it is strictly
worse. A mechanism is an algorithmic implementation of an
$\alpha$-approximation in undominated strategies~\cite{BLP09} if for every
outcome of the mechanism where every agent plays an undominated
strategy, the objective function value of the mechanism is within a
factor of $\alpha$ of the optimal, and every agent can easily compute
for any dominated strategy a strategy that dominates it.

\vspace{0.2in}
\begin{proofof}{Lemma~\ref{lem:undominated}}
Note that if agent $i$ desires only
one service $j\in J_i$, and refuses the service when offered, the
agent obtains a utility of $0$ regardless of others' strategies. On
the other hand, the strategy of accepting the service when offered has
strictly positive utility for some strategy profiles of others,
therefore it dominates the previous strategy.
\end{proofof}

\vspace{0.2in}

\begin{proofof}{Theorem~\ref{thm:mdmd-spm}}
We consider the matroid intersection setting first and assume that the
valuation distributions are regular. The non-regular case is
similar. Our SPM in this setting considers the hypothetical
single-dimensional instance $\Ifc$ defined in
Section~\ref{subsec:md-redn} and computes the probabilities $\probmi[j]$
with which Myerson's mechanism allocates the service $j$. We then set
$\probi[j]=\probmi[j]/2$ and
$\pricei=\disti[j]^{-1}(1-\probi[j])$. Note that for any $i$,
$\sum_{j\in J_i} \probi[j]\le 1/2$. Therefore, with probability at
least $1/2$, $i$ desires no service other than $j$ (we say that $j$ is
uniquely desired by $i$). Lemma~\ref{lem:undominated} shows that in this case, in any
undominated strategy implementation, if $i$ is offered $j$ and desires
it, then $i$ accepts $j$.

For any particular run of the mechanism, divide the set of all
services into three groups---$S$, the set of {\em sold} services, $B$
the set of services that are desired by their corresponding agents but
``blocked'' by services in $S$, and $U$ the set of services that are
desired by their corresponding agents and not in sets $S$ or $B$. Then
Lemma~\ref{lem:undominated} implies that services in $U$ are not uniquely
desired. Now, the expected total price in the union of the sets $S$,
$B$ and $U$ is exactly $\sum_j\pricei[j]\probi[j]$. Moreover, the
expected total price in $U$ is at most $1/2
\sum_j\pricei[j]\probi[j]$. Finally, following the proof of
Theorem~\ref{thm:matroidSpm2Approx}, the expected total price in $B$ conditioned on $S$ is
at most the total price contained in $S$. Therefore, putting
everything together we get that the expected total price obtained from
$S$ is at least $1/4\sum_j\pricei[j]\probi[j]$. By our choice of
$\prices$ and $\probs$, this is an $8$-approximation.

The argument for the combinatorial auction setting is identical and
based on Theorem~\ref{thm:bundles-spm}. We omit it for brevity.
\end{proofof}

\section{Approximating social welfare and other objectives via posted-price mechanisms}
\label{app:other-objectives}

We now show that our approach from Sections~\ref{sec:spm}
and \ref{sec:opm} in fact extends to the problem of maximizing any
objective that is linear in social value and revenue via SPMs.

We start with some definitions. For all $i\in [n]$ let
$g^i(\val,\price)=\alpha_i\val+\beta_i\price$ denote an arbitrary
linear function of $\val$ and $\price$. For a mechanism $\Alg$ with
payment rule $\prices$, let $G(\Alg,\prices)$ be the expected value of
$g$ over the outcome of the mechanism, that is, $G(\Alg,\prices) =
\expect_\vals[\sum_{i\in\Alg(\vals)}g^i(\vali,\pricei)]$. Define the
virtual value of $i$ with respect to $g^i$ to be 
\[\gvirti(\val)
= (\alpha_i+\beta_i)\val-\beta_i\frac{1-\disti(\val)}{\densi(\val)}\]
and the virtual surplus with respect to $G$ of a set $S$ of agents to
be $\gvs(S)=\sum_{i\in S}\gvirti(\vali)$. Then, the lemma below
follows from standard techniques, and allows us to ignore the payment
function in trying to maximize $G$.
\begin{lemma}
\label{lem:gvs}
For any truthful mechanism $\Alg$ with payment rule $\prices$, the
expected virtual surplus with respect to $G$ of $\Alg$ is equal to the
expected value of $G$ for $\Alg$'s outcome. That is,
\[G(\Alg,\prices) = \expect_\vals[\gvs(\Alg(\vals))]\]
\end{lemma}
The lemma suggests that a mechanism $\gMye$ with allocation rule
$\gMye(\vals) = \argmax_S\gvs(S)$ maximizes $G$ over the class of all
truthful mechanisms. However, as for revenue-maximizing mechanisms, in
order for this mechanism to be truthful, the distributions $\disti$
must satisfy a certain regularity condition.
\begin{definition}
\label{def:g-regular}
A one dimensional distribution distribution $\dist$ is regular with
respect to function $G$, if $\gvirt(\val)$ is monotone non-decreasing
in $\val$.
\end{definition}
The following theorem is straightforward:
\begin{theorem}
\label{thm:gmye}
If for all $i$, $\disti$ is regular with respect to $G$, the
mechanism $\gMye$ defined above is truthful and obtains the maximum
value of $G$ over the class of all truthful mechanisms.
\end{theorem}
In order to optimize $G$ over the class of SPMs in the matroid
setting, we follow an approach similar to the one in
Section~\ref{sec:spm}. Other approximations are similar. We focus on
the regular setting. Our approximately optimal mechanism is defined as
follows. Let $\gMye$ denote the optimal mechanism in
Theorem~\ref{thm:gmye} above. Let $\probgi$ denote the probability
that $\gMye$ serves agent $i$. Define for all $i$
\begin{alignat*}{2}
\probi & = \probgi,\\ 
\pricei & = \disti^{-1}(1-\probi), & \text{ and, }\\
\gi & = \left( \int_{\pricei}^{\infty} \gvirti(\vali) \densi(\vali)
d\vali \right) / \probi\\
\end{alignat*}
The SPM sets a price of $\pricei$ for agent $i$ and offers to serve
the agents in decreasing order of their corresponding $\gi$'s. The
$\gi$ reflects the expected virtual value we get from agent $i$ upon
serving the agent. We denote this mechanism by $\gspm$.

We first note that the performance of $\gspm$ can be bounded in terms
of the $\gi$'s. In particular, 
Lemma~\ref{lem:gvs} and the definition of $\gi$ imply that
\[G(\gspm) = \expect_{\vals}\left[\sum_{i\in \soldset} \gi\right]\]
where $\soldset$ is the set of agents that are allocated
service. Following the argument for
Theorem~\ref{thm:matroidSpm2Approx} we infer that since agents are
ordered in decreasing order of $\gi$,
$\expect_{\vals}\left[\sum_{i\in \soldset} \gi\right] \ge \frac
12 \sum_i \probi\gi$. In order to complete our argument, we bound the
performance of $\gMye$ in terms of the $\gi$'s.


\begin{lemma}
If for all $i$, $\disti$ is regular with respect to $G$, then
$G(\gMye)\le \sum_i \gi\probi$.
\end{lemma}
\begin{proof}
Let us consider the contribution of agent $i$ to the objective
function value for $\gMye$. This is no more than the objective
function value achieved by an optimal mechanism that sells only to $i$
and with probability at most $\probi$. By the definition of $\gvs$
and using regularity, this is exactly $\int_{\pricei}^{\infty}
\gvirti(\vali)\densi(\vali)d\vali$ where
$\pricei=\disti^{-1}(1-\probi)$. Finally, the integral is exactly
equal to $\gi\probi$ by the definition of $\gi$.
\end{proof}

We therefore have the following theorem:
\begin{theorem}
The mechanism $\gspm$ defined above obtains a $2$-approximation to the
objective $G$ in the matroid case when all the input distributions are
regular with respect to $G$.
\end{theorem}

Similar techniques prove analogues of other theorems in
Sections~\ref{sec:spm} and \ref{sec:opm} for arbitrary functions $G$.
Finally, we note that if the distributions are not regular as defined
in Definition~\ref{def:g-regular}, we can apply an ironing procedure
to the virtual values in much the same way as in Myerson's
approach. We leave the details to the reader.



\section{Revenue maximization through VCG mechanisms}
\label{sec:vcg}
\input{vcg.tex}

%% file: myerson-appendix.tex
\section{Myerson's mechanism and revenue bounds for truthful mechanisms}
\label{app:myerson}

Myerson's seminal work describes the revenue maximizing mechanism for
the Bayesian single-parameter mechanism design problem, BSMD,
described in Section~\ref{subsec:BOMD}. In Myerson's mechanism the
seller first computes so-called virtual values for each agent, and
then allocates to a feasible subset of agents that maximizes the
``virtual surplus''---the sum of the virtual values of agents in the
set minus the cost of serving that set of agents. These quantities are
formally defined as follows.

\begin{definition}
For a valuation $\vali$ drawn from $\disti$, the virtual valuation
of agent $i$ is given by
\begin{align*}
\virti(\vali) = \vali - \frac{1-\disti(\vali)}{\densi(\vali)}
\end{align*}
The virtual surplus of a set $S$ of agents is defined as
$\vsurp(S,\vals) = \sum_{i\in S} \virti(\vali)$. 
\end{definition}

\noindent
Myerson's optimal mechanism is based on the following observation.

\newcounter{temp}
\setcounter{temp}{\value{theorem}}
\setcounter{theorem}{\value{myeCount}}

\begin{proposition}
The expected revenue of any truthful single-parameter mechanism $M$ is
equal to its expected virtual surplus.
\end{proposition}
\setcounter{theorem}{\value{temp}}

A direct consequence of Proposition~\ref{lemma:Myerson} is that the
expected revenue maximizing mechanism would be one that maximizes
expected virtual surplus. Given a vector $\vals$ of values, Myerson's
mechanism serves the set $\argmax_S \vsurp(S,\vals)$. This mechanism
is truthful when the virtual valuation function is monotone
non-decreasing for every $i$, or in other words, the distribution
$\disti$ is {\em regular}. Note that we do not explicitly specify the
prices charged by the mechanism. These are uniquely determined by the
allocation rule assuming that agents that are not served pay nothing.

\begin{definition}
\label{def:regular}
A one dimensional distribution distribution $\dist$ is regular, if
$\virt(\val)$ is monotone non-decreasing in $\val$.
\end{definition}

\paragraph{Irregular distributions and ironed virtual values.}
When the distributions $\disti$ are irregular, that is,
Definition~\ref{def:regular} does not hold, Myerson's mechanism as
described above will no longer be truthful. Myerson fixed this case by
``ironing'' the virtual valuation function and converting it into a
monotone non-decreasing function. We skip the description of this
procedure; the reader is referred to \cite{BR-89, CHK-07} for
details. Let us denote the ironed virtual value of an agent with value
$\vali$ by $\ivirti(\vali)$. We then note the following.

\begin{proposition}\label{lemma:Ironing}
The expected revenue of any truthful single-parameter mechanism $M$ is
no more than its expected ironed virtual surplus. If the probability
with which the mechanism serves agent $i$, as a function of $\vali$,
is constant over any valuation range in which the ironed virtual value
of $i$ is constant, the expected revenue is equal to expected ironed
virtual surplus.
\end{proposition}

Myerson's mechanism serves a subset of agents that maximizes the
ironed virtual surplus, breaking ties in an arbitrary but consistent
manner. Denoting the revenue of a mechanism $\A$ by $\RevAlg$ and the
revenue of Myerson's mechanism $\Mye$ by $\RevMye$ we get the
following:

\begin{theorem}\label{theorem:Ironing}
$\RevMye \geq \RevAlg$ for every truthful mechanism $\A$.
\end{theorem}

\paragraph{Bounding the revenue of the Bayesian optimal mechanism.}
Propositions~\ref{lemma:Myerson} and \ref{lemma:Ironing} give one
approach of bounding the expected revenue of Myerson's mechanism. We
now develop a different bound that is useful in proving performance
guarantees for posted-price mechanisms.
\setcounter{temp}{\value{theorem}}
\setcounter{theorem}{\value{myeVsSumPqCount}}
\begin{lemma}
If $\disti$ is regular for each $i$, for any truthful mechanism $M$
over the $n$ agents, the revenue of $M$ is bounded from above by
$\sum_i \pricemi \probmi$ where $\probmi$ is the probability (over
$\vali[1],\cdots,\vali[n]$) with which $M$ allocates to agent $i$ and
$\pricemi=\disti^{-1}(1-\probmi)$.

Furthermore for every $i$ (with a regular or non-regular value
distribution), there exist two prices $\underline{\pricei}$ and
$\overline{\pricei}$ with corresponding probabilities
$\underline{\probi}$ and $\overline{\probi}$, and a number $x_i\le 1$,
such that $x_i\underline{\probi} + (1-x_i)\overline{\probi} =
\probmi$, and the expected revenue of $M$ is no more than $\sum_i
x_i\underline{\pricei}\underline{\probi} +
(1-x_i)\overline{\pricei}\overline{\probi}$.
\end{lemma}
\setcounter{theorem}{\value{temp}}

\begin{proof}
We prove the regular case first. Consider the revenue that $M$ draws
from serving agent $i$. This is clearly bounded above by the optimal
mechanism that sells to only $i$, but with probability at most
$\probmi$. By Proposition~\ref{lemma:Myerson}, such a mechanism should
sell to agent $i$ with probability $1$ whenever the value of the agent
is above $\disti^{-1}(1-\probmi)$ and with probability $0$
otherwise. The revenue of the optimal such mechanism is therefore
$\pricemi\probmi$.




In the non-regular case, note that the value $\pricemi$ may fall in a
valuation range that has constant ironed virtual value. Let
$\underline{\pricei}$ denote the infimum $\inf\{\val:\ivirti(\val) =
\ivirti(\pricemi)\}$ of this range and $\overline{\pricei}$ denote the
supremum $\sup\{\val:\ivirti(\val) = \ivirti(\pricemi)\}$. Let
$\underline{\probi}=1 - \disti(\underline{\pricei})$ and
$\overline{\probi}=1 - \disti(\overline{\pricei})$. Then,
$\overline{\probi}\le\probmi \le \underline{\probi}$, and there exists
an $x_i$ such that $x_i\underline{\probi} + (1-x_i)\overline{\probi} =
\probmi$. Now an easy consequence of Proposition~\ref{lemma:Ironing} is
that the optimal mechanism with selling probability $\probmi$ sells to
the agent with probability $x_i$ if the agent's value is between
$\underline{\pricei}$ and $\overline{\pricei}$, and with probability
$1$ if the value is above $\overline{\pricei}$. The revenue of this
mechanism is exactly $x_i\underline{\probi}\underline{\pricei} +
(1-x_i)\overline{\probi}\overline{\pricei}$.
\end{proof}

%% file: spm-lb.tex
Blumrosen and Holenstein~\cite{BH-08} give an example where the gap
between the revenue of the optimal SPM and that of Myerson's mechanism
is a factor of $\sqrt{\pi/2}\approx 1.253$. We reproduce the example
here for completeness. There are $n$ agents, each with a value
distributed independently according to function $\dist(\val) =
1-1/\val^2$. The seller has one item to sell. Then, the expected
revenue of Myerson's mechanism is $\Gamma(1/2) \sqrt{n}/2$, where
$\Gamma()$ is the Gamma function. On the other hand, the expected
revenue of the optimal SPM can be computed to be
$\sqrt{n/2}$. Therefore, we get a gap of $\Gamma(1/2)/\sqrt{2} =
\sqrt{\pi/2}\approx 1.253$.

%% file: uniform-approx.tex

 We first prove Theorem~\ref{thm:uniform-spm}
 for 1-uniform matroids. The Revenue $\Revp[\spm]$ of the SPM described in 
 Section~\ref{sec:spm} 
 can be written as
 \begin{align*}
 \Revp[\spm] &= \sum_{i=1}^{n} \offerProb\pricemi\probmi
 = \sum_{i=1}^{n}\prod_{j=1}^{i-1}(1-\probi[j])\pricemi\probmi, 
 \end{align*}
 where $\offerProb = \prod_{j=1}^{i-1}(1-\probi[j])$ is the 
 probability that agent $i$ is offered service. Note
 that $\offerProb \geq \offerProb[j]$ for $i \leq j$.
Let $p$ be the price satisfying the equation   
\begin{equation}\label{eqn:samePrice}
\sum_{i}\pricei\probi = p\sum_{i}\probi.
\end{equation}
 We now prove that 
 among the set of all product 
 distributions $G = (G_1 \times G_2\times \dots \times G_n)$
 which satisfy
\begin{itemize}
\item Pr[Myerson's mechanism serves agent $i$]  = $\probi$; and
\item $\sum_{i}G_i^{-1}(1-\probi)\probi = \sum_{i}\pricei\probi$,
\end{itemize}
the revenue obtained is lowest when $G$ is the distribution for which
all the prices are equal, i.e. $G_i^{-1}(1-\probi) = p$ for all agents  $i$. 
Let $\Revp[\spm]_{eq}$ denote the revenue of the SPM whose
prices are equal to $p$ for all agents.
 \begin{lemma}\label{lemma:samePriceIsBad}
 It is always the case that $\Revp[\spm] \geq \Revp[\spm]_{eq}$.
 \end{lemma}
 \begin{proof}
 Let $\delta_i = \probi(\pricei-p)$. So we have
 \begin{align*}
 \Revp[\spm] &= \sum_{i=1}^{n} \offerProb\pricei\probi = \sum_{i=1}^{n}\offerProb
  (p\probi + \delta_i) = \Revp[\spm]_{eq} + \sum_{i=1}^{n}\offerProb\delta_i \geq \Revp[\spm]_{eq},
 \end{align*}
 where the inequality follows from observing that:
 \begin{itemize}
 \item $\offerProb$'s are in descending order;
 \item $\exists j$ such that $\delta_i$ is non-negative for all $i \leq j$ and negative otherwise; and
 \item $\sum_{i}\delta_i = 0$ (By~\eqref{eqn:samePrice}) .
 \end{itemize}
\end{proof}
 \begin{theorem}\label{thm:1uniformSPM}
 The revenue $\Revp[\spm]$ of the SPM $\spm$ is a $\frac{e}{e-1}$ approximation to the
 expected revenue of Myerson's mechanism in the case of a $1$-uniform
  matroid.
 \end{theorem}
 \begin{proof}
 Let $\sum_{i}\probi = s$ ($\leq$ 1).
 Lemma~\ref{lemma:samePriceIsBad} implies the theorem when
 $\Revp[\spm]_{eq} \geq( 1-\frac{1}{e})ps$.  We can see this holds, since
 \begin{align}
 \notag
   \Revp[\spm]_{eq} &= p(\text{Pr[Some agent is served]}) = p(1-\text{Pr[No agent is served]})\\
   \notag 
&= p\left(1-\prod_{i=1}^{n}(1-\probi)\right) \\
 \label{eqn:maxProdWhenEqual}
 &\geq p\left(1-\prod_{i=1}^{n}(1-s/n)\right) \\
 \notag
 &\geq(1-1/e)ps,
 \end{align}
 where~\eqref{eqn:maxProdWhenEqual} follows since the product is maximized when the $\probi$'s are all equal.
 \end{proof}

\noindent
Next we consider the $k$-uniform case.

 \begin{theorem}\label{thm:kuniformSPM}
 The revenue $\Revp[\spm]$ of the SPM $\spm$ is a $\frac{e}{e-1}$ approximation to the
 expected revenue of Myerson's mechanism in the case of a $k$-uniform matroid.
\end{theorem}
\begin{proof}
Our proof technique is closely related to the proof for 1-uniform
matroids. If we define $\Revp[\spm]_{eq}$ as defined above for the
$1$-uniform case, then the proof of Lemma~\ref{lemma:samePriceIsBad}
extends to $k$-uniform matroids also. Thus it would be enough to argue
that $\Revp[\spm]_{eq}$ achieves a $e/(e-1)$ approximation.  Let $p$
be the common price for all agents which satisfies~\eqref{eqn:samePrice}.

For any set of probabilities $\{\probi\}$ in the k-item case, let us define
$\probi' = \probi/k$. Note that the probabilities $\{\probi'\}$ form 
a valid set of probabilities for a 1-item case because
\begin{align*}
\sum_{i}\probi' = \sum_{i}\probi/k \leq 1
\end{align*}
Let $\offerProb'$ denote the probability that agent $i$ is 
considered for service in the 1-item case. We can come up with distributions
$\disti'$ for the 1-item case such that the price $\disti'^{-1}(1-\probi')$ is the same
for all agents and is equal to $p$. By Theorem~\ref{thm:1uniformSPM}, we know that the 
revenue in this 1-item case is at least $(1-1/e)\sum_{i} p\probi'$.
We prove that we get the same approximation factor of $e/(e-1)$ for the $k$-item
case by the following induction. We assume that for $j-1 \leq n$, the 
revenue $R_{j-1}$ from the first $j-1$ items is at least $k$ times the 
revenue $R_{j-1}'$ from the first $j-1$ items in the corresponding 1-item case
i.e. $\sum_{i=1}^{j-1}\offerProb p\probi \geq k\cdot\sum_{i=1}^{j-1}\offerProb'p\probi'$.
We prove the same for $j$ through two cases.
 \begin{enumerate}
\item  If $\offerProb[j] \geq \offerProb[j]'$, then we are done, because we know
that revenue $R_j$ from the first $j$ items can be written as 
\begin{align*}
 R_j = R_{j-1} + \offerProb[j]p\probi[j] \geq k(R_{j-1}') + k\offerProb[j]'p\probi[j]' = kR_j'.
\end{align*}
 The inequality uses the induction hypothesis.
\item If $\offerProb[j] < \offerProb[j]'$, we show that the revenue obtained is better
  than when $\offerProb[j] = \offerProb[j]'$ and then we will be done. To see this observe
  that the revenue $R_{j}$ can be written as being conditioned on whether
  or not $k$ items were sold in the first $j-1$ items. So we have
 \begin{align*}
 R_j &=(1-\offerProb[j])kp + \offerProb[j]\big(p\probi[j] + \text{ E[ Revenue from first $j-1$ $|$ at most $k-1$ of first $j-1$ are served ]}\big);
 \intertext{since}
 kp &\geq \big(p\probi[j] + \text{E[Revenue from first $j-1$ $|$ at most $k-1$ of first $j-1$ are served]}\big),
 \end{align*}
 the revenue only decreases by increasing $\offerProb[j]$ to $\offerProb[j]'$.
\end{enumerate}
 
Thus in either case, the $k$-item case has a better revenue, guaranteeing
an approximation factor of $\frac{e}{e-1}$.
 \end{proof}
\begin{corollary}
 The revenue $\Revp[\spm]$ of the SPM $\spm$ is a $\frac{e}{e-1}$ approximation to the
 expected revenue of Myerson's mechanism in the case of a partition-matroid.
\end{corollary}

\input{spm-lb-2.tex}

%% file: spm-lb-2.tex
We note that this analysis is tight. In particular, consider an
example with $n$ \iid agents and a seller with one item. Suppose that
the value of each agent is independently $1$ with probability
$1-\epsilon$ and $0$ otherwise for some small $\epsilon>0$. Then, the
expected revenue of Myerson's mechanism is equal to the probability
that at least one agent has value $1$, which is $1-o(1)$. The
probability with which Myerson's mechanism serves a particular agent
is $(1-o(1))/n$. Therefore, our mechanism sets a price of $1$ for each
agent, and offers the price to each agent with probability roughly
$1/n$ until some agent accepts. So the revenue of our mechanism is
roughly $1-(1-1/n)^n = 1-1/e$. A similar gap can be obtained even if
the SPM decides to offer a price to the last agent with certainty when
no other agent accepts the item. Note that there is a simple SPM that
in this case obtains the optimal revenue---offer a price of $1$ to
each agent in turn with certainty until the item is sold.

%% file: prophet.tex
\newcommand{\thresholdi}[1][i]{{t}_{#1}}
\newcommand{\highThresh}{{\highThreshVar}^{\ast}}
\newcommand{\lowThresh}{{\lowThreshVar}^{\ast}}
\newcommand{\highThreshVar}{b}
\newcommand{\lowThreshVar}{a}
\newcommand{\pos}[1]{{\left({#1}\right)}^{+}}
\newcommand{\prophetRVi}[1][i]{X_{#1}}
\newcommand{\prophetRVoi}[1][i]{X_{(#1)}}
\newcommand{\indicator}[1]{I\left({#1}\right)}


Consider the following setting from \cite{Cahn84}: a gambler is
presented samples from $n$ distributions in order,
$\prophetRVi[1],\dots,\prophetRVi[n]$. For each sample, the gambler
must decide whether to pick this sample (and end the game) or skip it
(to never return to it). The gambler can choose at most one of the
samples, and obtains a reward equal to the value of the sample. Can
the gambler do nearly as well in expectation as a prophet that knows
the maximum value in the sample? Samuel-Cahn \cite{Cahn84} shows that
there is a simple threshold rule for picking samples that uses a
common threshold for each random variable, such that the expected
value of the gambler is within a factor of $2$ of the expected value
obtained by the prophet. We first extend this result of \cite{Cahn84}
to the case where both gambler and prophet can pick $k$ values, and
then describe how it applies to our setting of maximizing revenue.


We begin with some definitions.  Given a collection of $n$
independent, nonnegative random variables
$\prophetRVi[1],\dots,\prophetRVi[n]$, we consider extending the
prophet inequalities to the case where the gambler and the prophet are
each allowed $k$ choices.  Let
$\prophetRVoi[1]\ge\dots\ge\prophetRVoi[n]$ be the order statistics
for $\prophetRVi[1],\dots,\prophetRVi[n]$.  For a value $x$, let
$\pos{x}$ denote the positive portion of $x$,
i.e. $\pos{x}=\max(0,x)$.

For a constant $c$, let $\thresholdi[1](c),\dots,\thresholdi[k](c)$
denote the $k$ indices selected by a threshold stopping rule using
$c$, i.e. $\thresholdi(c)$ is the lesser $(n-k+i)$ and the
$i^\text{th}$ smallest index $j$ such that $\prophetRVi[j]{\ge}c$ (or
simply the former when the latter does not exist).

Let $\lowThresh$ and $\highThresh$ be the unique solutions to the
equations
\begin{equation*}
  \lowThreshVar = \sum_{i=1}^{k}\expect\pos{\prophetRVoi-\lowThreshVar/k}
  ,\qquad\text{and}\qquad
  \highThreshVar = \sum_{i=1}^{n}\expect\pos{\prophetRVi-\highThreshVar/k},
\end{equation*}
respectively.  Then it must be the case that
$\lowThresh\le\highThresh$, and we get the following theorem.

\begin{theorem}
\label{thm:kChoiceProphet}
  For $\lowThresh\le k\cdot c \le\highThresh$, we have that
  $\sum_{i=1}^{k}\expect\left[\prophetRVoi\right]\le2\sum_{i=1}^{k}\expect\left[\prophetRVi[\thresholdi(c)]\right]$.
\end{theorem}
\begin{proof}
First, we note that for any threshold $t$
\begin{equation*}
  \sum_{i=1}^{k}\expect\left[\prophetRVoi\right]
  \le \sum_{i=1}^{k}\expect\left[t + \pos{\prophetRVoi-t}\right]
  = k\cdot t + \sum_{i=1}^{k}\expect\pos{\prophetRVoi-t},
\end{equation*}
which implies
$\sum_{i=1}^{k}\expect\left[\prophetRVoi\right]\le2\lowThresh$ with
the substitution $t=\lowThresh/k$.  Now, any time
$\thresholdi[k](c)<n$, we know there are at least $k$ $\prophetRVi$ at
or above our threshold $c$, and so
\begin{equation*}
  \sum_{i=1}^{k}\expect\left[\prophetRVi[\thresholdi(c)]\right] 
  \ge
  kc\cdot\Pr[\thresholdi[k]<n] + \sum_{i=1}^{k}\expect\pos{\prophetRVi[\thresholdi(c)]-c}.
\end{equation*}
Let $\indicator{\mathcal{E}}$ denote the indicator random variable for
event $\mathcal{E}$.  Considering the second term above, we see that
\begin{align*}
  \sum_{i=1}^{k}\expect\pos{\prophetRVi[\thresholdi(c)]-c}
  &=\sum_{i=1}^{k}\sum_{j=1}^{n}\expect\left[\pos{\prophetRVi[j]-c}\indicator{\thresholdi(c)=j}\right]\\
  &=\sum_{j=1}^{n}\expect\left[\pos{\prophetRVi[j]-c}\indicator{\thresholdi[k](c)>j-1}\right]\\
&\ge \sum_{j=1}^{n}\expect\pos{\prophetRVi[j]-\highThresh/k}\cdot\Pr[\thresholdi[k](c)>n-1].
\end{align*}
Since the sum in the last line is precisely $\highThresh$, we
  see our choice of $c$ gives
  $\sum_{i=1}^{k}\expect\left[\prophetRVi[\thresholdi(c)]\right] \ge
  kc \ge \lowThresh$ as claimed.
\end{proof}

We now have the necessary results in place to proceed with the proof
of Theorem~\ref{thm:uniformOpmApprox}.

%% file: uniform-opm.tex
\begin{proofof}{Theorem~\ref{thm:uniformOpmApprox}}
  We prove our revenue bound via virtual values.  We assume that all
  the $\disti$ are regular.  In fact, we see that we choose prices
  which do not distinguish (for a given agent) between differing
  valuations that yield the same virtual value, and so by Proposition~\ref{lemma:Ironing}
  may use ironed
  virtual values in the irregular case to
  achieve the same result.

We cannot immediately apply Theorem~\ref{thm:kChoiceProphet}, since
virtual values can, in general, be negative; so we consider
$\pos{\virti}$ in place of $\virti$. (Note that Myerson's mechanism
never selects an agent with a negative virtual value, and neither will
our mechanism.)

Let $c$ be the threshold from applying
Theorem~\ref{thm:kChoiceProphet} to the random variables
$\pos{\virti}$, and $\pricei=\inf\{v:\virti(v) \ge c\}$. 
Then, we can see that the expected revenue of an OPM
using these prices is
\begin{equation*}
  \RevObl
  =\expect\left[\sum_{i=1}^{k}\pos{\virt_{\thresholdi(c)}}\right]
  \ge\frac{1}{2}\expect\left[\sum_{i=1}^{k}\pos{\virt_{(i)}}\right]
  =\RevMye/2.
\end{equation*}
\end{proofof}

%% file: opm-lb.tex
\paragraph{An example with a gap of $2$.} We now show that OPMs cannot
approximate the optimal revenue to within a factor better than $2$
even in the single-item setting. Consider a seller with one item and
two agents. The first agent has a fixed value of $1$. The second has a
value of $1/\epsilon$ with probability $\epsilon$ and $0$ otherwise,
for some small constant $\epsilon>0$. Then, the optimal mechanism can
obtain a revenue of $2-\epsilon$ by first offering a price of
$1/\epsilon$ to the second agent, and then a price of $1$ to the first
if the second declines the item. On the other hand, if the mechanism
is forced to offer the item to the first agent first, then it has two
choices: (1) offer the item at price $1$ to agent 1; the agent always
accepts, and (2) skip agent $1$ and offer the item at price
$1/\epsilon$ to agent 2; the agent accepts with probability
$\epsilon$. In either case, the expected revenue of the mechanism is
$1$.

%% file: graphical-opm.tex
\newcommand{\incident}{\delta}
\newcommand{\probEdge}{\prob_e}
\newcommand{\probVert}{\prob_v}

\begin{proofof}{Theorem~\ref{thm:graphicalOpmApprox}}
Our technique here is to partition the elements of the matroid such
that we may treat each part as a $1$-uniform matroid yet still respect
the original feasibility constraint, and achieve good revenue while
doing so.

  Let $G=(V,[n])$ be the graph defining our matroid constraint.  Let
  $\incident(v)$ denote the set of edges incident on a vertex $v$, and
  for each $v\in V$ define
  $\probVert=\sum_{i\in\incident(v)}\probi$.  Now, we can see that
  \begin{equation*} 
  \sum_{v \in V} \probVert = \sum_{i \in E}2\probi \le 2(\abs{V}-1), 
  \end{equation*} 
  which can only hold
  if there exists $v$ such that $\probVert\le2$; let $\incident(v)$ be
  one of our partitions.  Furthermore, the edge set $\incident(v)$
  forms a cut in $G$, and so given an independent set of edges from
  $E-\incident(v)$ we may add any single edge from $\incident(v)$ while
  retaining independence.  We apply this argument recursively to
  $(V-{v},E-\incident(v))$ to form the rest of our partition.  At the
  end, we have a partition of $E$ such that each part has total mass
  no more than $2$, and any collection of edges using no more than one
  edge from each part is independent.
  
  We now show that within each part $P$, we can achieve expected
  revenue at least a third of what Myerson's mechanism received from
  that part, via an application of Theorem~\ref{thm:kChoiceProphet}.
  Note that the revenue achieved by offering an agent $i$ a price of
  $p_i$ is a random variable, and these random variables are
  nonnegative and independent.  Furthermore, we can successfully apply
  a threshold rule to them -- we only offer to an agent if $p_i$ is
  above our threshold, and they accept if and only if our stopping
  rule would have chosen this agent.

  Let $p=\sum_{i\in P}q_i\pos{p_i-p}$.  Then $p$ is precisely the
  upper bound on acceptable thresholds for
  Theorem~\ref{thm:kChoiceProphet} applied to our specified random
  variables, allowing one choice.  From the proof of that theorem, we
  can see that applying the threshold $p$ results in an expected
  revenue of at least $p$; on the other hand, 
  \begin{equation*}
  \sum_{i \in P}p_iq_i \le \sum_{i \in P}q_i(p + \pos{p_i -p}) = p(1 +
  \sum_{i \in P}q_i) \le 3p.  
  \end{equation*}
\end{proofof}

%% file: matroidInt-opm.tex
\begin{proofof}{Theorem~\ref{thm:opmPartitionMatroidIntersection}}
We describe the mechanism which achieves a $6.75$-approximation when the distributions are
regular. Appendix ~\ref{app:non-regular} sketches the extension to the non-regular case.


Let $\probi = \probmi/3$ and $\pricei = \disti^{-1}(1-\probi)$. Note that $\pricei \geq \pricemi$.
The mechanism 
serves agents in any arbitrary order, but offers a price of $\pricei$ for agent $i$.

Let $\offerProb$ denote the probability that agent $i$ is considered for service. We
prove that $\offerProb \geq 4/9$ for all $i$. This would prove the theorem,  as
the expected revenue is 
\begin{align*}
\RevObl = \sum_{i}\offerProb\pricei\probi \geq \sum_{i}(4/9)\pricemi(\probmi/3) \geq \sum_{i}(1/6.75)\pricemi\probmi.
\end{align*}
Let $\mathcal{M}_1,\ \mathcal{M}_2$ be the two partition matroids.
Let agent $i$ be in  partition $P_1$ of $\mathcal{M}_1$ and 
in partition $P_2$ of $\mathcal{M}_2$.
Let $k_1$ be the maximum number of elements in $P_1$ that can be present in an 
independent set of $\mathcal{M}_1$ and let $k_2$ be the maximum number of elements
in $P_2$ that can be present in an independent set of $\mathcal{M}_2$. We then have that for $j=1,2$ that
the expected number of agents in $P_j$ desiring service is
\begin{align*}
\sum_{i \in P_j}\probi 
\leq k_j/3. 
\end{align*}
Define $\mathcal{E}_j$ to be the event that
at most $k_j-1$ agents from $P_j$ desire service 
for $j=1,2$; then 
agent $i$ is always considered for service when events $\mathcal{E}_1$ and $\mathcal{E}_2$ both happen.
By Markov's inequality, it is clear that $\Pr[\bar{\mathcal{E}_1}] \leq 1/3$ and 
$\Pr[\bar{\mathcal{E}_2}] \leq 1/3$.  So we may conclude that
\begin{align*}
  \offerProb \geq \Pr[\mathcal{E}_1 \cap \mathcal{E}_2] 
  = \Pr[\mathcal{E}_1]\cdot \Pr[\mathcal{E}_2 | \mathcal{E}_1] 
  \geq (2/3)\cdot (2/3),
\end{align*}
and  the claim follows.
\end{proofof}

%% file: vcg.tex
A consequence of our constant-factor approximation to revenue through
SPMs is that in matroid settings VCG mechanisms with appropriate
reserve prices are near-optimal in terms of revenue. This follows from
noting, as we show below, that VCG mechanisms perform no worse in
terms of expected revenue than SPMs with the same reserve
prices. Although VCG mechanisms aim to maximize the social welfare of
the outcome, setting high enough reserve prices allows them to also
obtain good revenue.

Formally, a Vickrey-Clarke-Groves (VCG) mechanism $\vcg[\prices]$ with
reserve prices $\prices$ serves the set $S$ of agents, with
$\vali\ge\pricei$ for all $i\in S$, that maximizes $\sum_{i\in
S} \vali$.

Hartline and Roughgarden~\cite{HR-09} show that in several
single-parameter settings the VCG mechanism with monopoly reserve
prices gives a constant factor approximation to revenue. This result
holds when all the value distributions satisfy the so-called monotone
hazard rate condition, or with a matroid feasibility constraint when
all the value distributions are regular. Their result does not extend
to the case of matroids with general (non-regular) value
distributions. One of the main questions left open by their work is
whether there is some set of reserve prices (not necessarily equal to
the monopoly reserve prices) for which the VCG mechanism gives a
constant factor approximation to revenue in the matroid setting with
general value distributions. We answer this question in the positive.
We use the following fact about matroids.

\begin{proposition}
\label{prop:matroid-bijection}
Let $B_1$ and $B_2$ be any two independent sets of equal size in a
matroid set system $\sets$. Then there is a bijective function
$g:B_1\setminus B_2\rightarrow B_2\setminus B_1$ such that for all
$e\in B_1\setminus B_2$, $B_1\setminus\{e\}\cup\{g(e)\}$ is
independent in $\sets$.
\end{proposition}

\begin{theorem}
\label{thm:vcg-matroid}
For any instance of the single-parameter Bayesian mechanism design
problem with a matroid feasibility constraint, there exists a set of
reserve prices such that the expected revenue of the VCG mechanism
with those reserve prices is at least half of the expected revenue of
Myerson's mechanism.
\end{theorem}

\begin{proof}
We prove that when the set system $\sets$ is a matroid, for any
collection of prices $\prices$, the revenue of the SPM $\spm[\prices]$
is no more than the revenue of the VCG mechanism $\vcg[\prices]$. The
result then follows from Theorem~\ref{thm:matroidSpm2Approx}.

Fix a value vector $\vals$ and let $A$ denote the set served by
$\spm[\prices]$ and $B$ denote the set served by
$\vcg[\prices]$. Then, since both mechanisms serve a maximal
independent set among the set of agents with $\vali\ge\pricei$, we
have $|A|=|B|$. Proposition~\ref{prop:matroid-bijection} then implies
the existence of a bijection $g$ such that for all $e\in B\setminus
A$, $B\setminus\{e\}\cup\{g(e)\}$ is independent. This implies that
$\vcg[\prices]$ charges $e$ a price of at least the value of $g(e)$,
which is at least the reserve price $\pricei[g(e)]$. On the other
hand, by definition, the price charged to any $e\in B\cap A$ is at
least $\pricei[e]$. Therefore, the revenue of $\vcg[\prices]$ in this
case is at least $\sum_{e\in B\cap A} \pricei[e] + \sum_{e\in
  B\setminus A} \pricei[g(e)] = \sum_{e\in A} \pricei[e]$ which is
equal to the revenue of $\spm[\prices]$.
\end{proof}